\documentclass[aps, twocolumn, pra, amsmath, superscriptaddress,preprintnumbers]{revtex4-1}
\UseRawInputEncoding
\usepackage{graphicx}
\usepackage{dcolumn}
\usepackage{bm}
\usepackage{amsmath}
\usepackage{subfigure}
\usepackage{amsfonts}
\usepackage{xcolor}
\usepackage{appendix}
\usepackage{multirow}
\usepackage{booktabs}
\usepackage{tabularx}
\usepackage[colorlinks,
            linkcolor=blue,
            anchorcolor=red,
            citecolor=blue
            ]{hyperref}
\usepackage{amsthm}
\usepackage{setspace}
\usepackage{lineno}
\usepackage{footnote}

\usepackage[draft]{changes} %%%%---annotations are visible
\begin{document}
\title{Quantum control of bosonic modes with superconducting circuits}
\author{Wen-Long Ma}
%\email{wenlongma@semi.ac.cn}
\affiliation{State Key Laboratory for Superlattices and Microstructures, Institute of Semiconductors, Chinese Academy of Sciences, Beijing 100083, China}
\affiliation{Pritzker School of Molecular Engineering, University of Chicago, Illinois 60637, USA}
\author{Shruti Puri}
\affiliation{Department of Applied Physics and Physics, Yale University, New Haven, Connecticut 06511, USA}
\affiliation{Yale Quantum Institute, Yale University, New Haven, Connecticut 06511, USA}
\author{Robert J. Schoelkopf}
\affiliation{Department of Applied Physics and Physics, Yale University, New Haven, Connecticut 06511, USA}
\affiliation{Yale Quantum Institute, Yale University, New Haven, Connecticut 06511, USA}
\author{Michel H. Devoret}
\affiliation{Department of Applied Physics and Physics, Yale University, New Haven, Connecticut 06511, USA}
\affiliation{Yale Quantum Institute, Yale University, New Haven, Connecticut 06511, USA}
\author{S. M. Girvin}
\affiliation{Department of Applied Physics and Physics, Yale University, New Haven, Connecticut 06511, USA}
\affiliation{Yale Quantum Institute, Yale University, New Haven, Connecticut 06511, USA}
\author{Liang Jiang}
\email{liang.jiang@uchicago.edu}
\affiliation{Pritzker School of Molecular Engineering, University of Chicago, Illinois 60637, USA}

\date{\today }

\begin{abstract}
Bosonic modes have wide applications in various quantum technologies, such as optical photons for quantum communication, magnons in spin ensembles for quantum information storage and mechanical modes for reversible microwave-to-optical quantum transduction. There is emerging interest in utilizing bosonic modes for quantum information processing, with circuit quantum electrodynamics (circuit QED) as one of the leading architectures. Quantum information can be encoded into subspaces of a bosonic superconducting cavity mode with long coherence time. However, standard Gaussian operations (e.g., beam splitting and two-mode squeezing) are insufficient for universal quantum computing. The major challenge is to introduce additional nonlinear control beyond Gaussian operations without adding significant bosonic loss or decoherence. Here we review recent advances in universal control of a single bosonic code with superconducting circuits, including unitary control, quantum feedback control, driven-dissipative control and holonomic dissipative control. Various approaches to entangling different bosonic modes are also discussed.

\end{abstract}

%Received: 2021/2/21

%Revised: 2021/5/20

%Accepted: 2021/5/24

\maketitle

\section{Introduction}
Quantum computation holds the promise of solving some specific problems, such as factorization of large integers and simulation of quantum many-body problems \cite{Nielsen2000}, much faster than any known classical computer. To build such a quantum computer, the physical platform should work in the quantum regime with long coherence time, fast quantum operations and good scalability, which are daunting obstacles for current technologies. The promising strategies to overcome such obstacles are quantum error correction (QEC) \cite{Shor1995,Knill1997,Lidar2013} and fault-tolerant (FT) quantum computation \cite{Preskill1998}, where the coherence time of the quantum memories can be extended and the quantum operations can tolerate some low-probability errors (including errors in the QEC circuit) below a certain threshold.

In the prototypical model for quantum computation - the quantum circuit model, a quantum bit of information (qubit) is encoded into a two-level system, called a physical qubit, and the usual approach for QEC is to encode a logical qubit into some subspace of multiple physical qubits, so that different error processes lead to distinguishable syndromes and can therefore be corrected. However, the increased number of physical qubits for a logical qubit introduces more decoherence for the system to correct. Moreover, the logical gate operations become quite complicated since multiple physical systems need to be addressed simultaneously. Hence, it is still an outstanding experimental challenge to build a more robust quantum register using multiple physical qubits.

An alternative scheme is to encode the quantum information into bosonic modes such as harmonic
oscillators \cite{Braunstein2005,Weedbrook2012}. A single bosonic mode already provides an infinitely
large Hilbert space, from which we choose a logical subspace for an error-correcting
code \cite{Chuang1997,Braunstein1998,Gottesman2001,Cochrane1999,Michael2016,Albert2018b}.
Such bosonic QEC modes can be hardware-efficient compared to the conventional QEC codes based
on multiple qubits. Moreover, the bosonic modes often have relatively simple decoherece processes
(mainly bosonic excitation loss channel) during which the bosonic excitations are lost one by
one \cite{Blais2021}. There have been several error-correcting encoding schemes in
a single bosonic mode proposed to date, including the Gottesman-Kitaev-Preskill (GKP)
codes \cite{Gottesman2001,Noh2018,Royer2020}, cat codes \cite{Cochrane1999,Li2017,Bergmann2016},
binomial codes \cite{Michael2016}, rotation-symmetric codes \cite{Grimsmo2020} and other variations
 \cite{Albert2018b,Li2019}. The GKP codes, consisting of superpositions of highly squeezed states,
 are not only protected against small shifts in position but also have been shown to perform well
 against the more realistic amplitude damping channel \cite{Albert2018b}. The cat codes use
 superpositions of coherent states evenly distributed around a circle in phase space, which
  can be protected against (single or multiple) bosonic excitation loss and dephasing errors.
  The binomial codes exploit superpositions of Fock states weighted with binomial coefficients,
  which can exactly correct the bosonic excitation loss, gain and dephasing errors up to a specific degree. %\deleted{Recently QEC based on cat codes or binomial codes in superconducting cavities have reached or approached the break-even point

\begin{figure}
\includegraphics[width=3.5in]{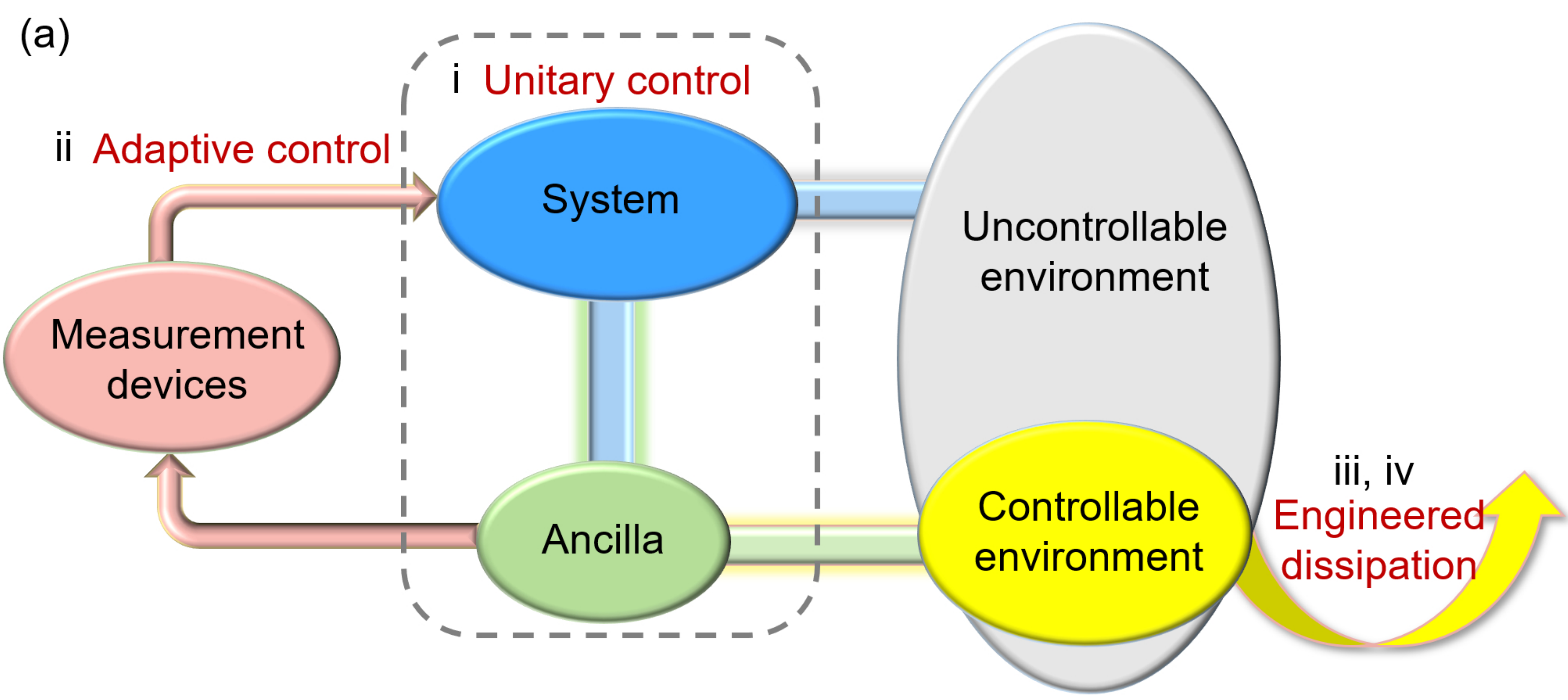}
\includegraphics[width=3.5in]{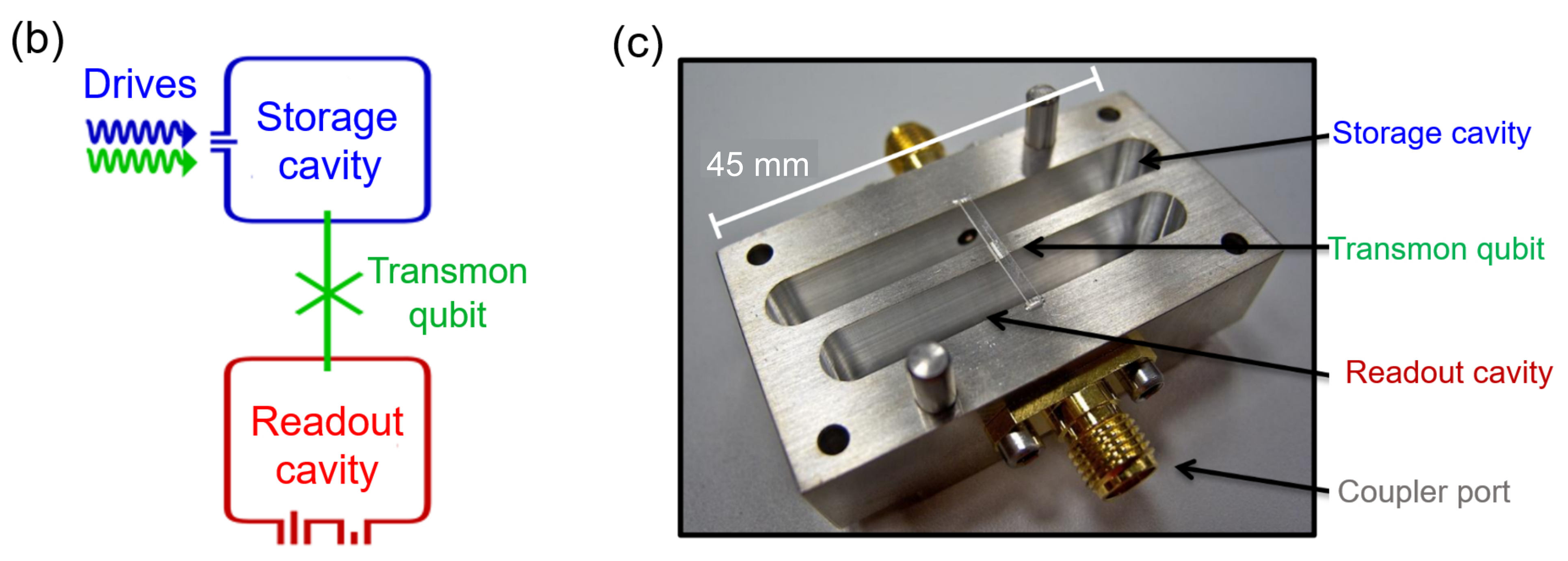}
\caption{(Color online) (a) Schematic of various approaches for controlling a quantum system: (i) unitary control on the system alone or both the system and an ancilla; (ii) quantum feedback control based on measurement of the ancilla; (iii) driven-dissipative control with either engineered dissipation or Hamiltonian engineering; (iv) holonomic quantum control based on only engineered dissipation. (b), (c) Schematic and device photograph of a circuit QED system modeled as a coupled qubit-oscillator system. The storage cavity with long coherence time is used to encode quantum information, the transmon qubit acts as an ancilla for universal control of the storage cavity, and the readout cavity with short coherence time is used for qubit readout. Both the storage cavity and transmon qubit can be addressed by microwave control fields. Reprinted with permission from Refs. \cite{Shen2017,Kirchmair2013}. }
\label{control}
\end{figure}

For bosonic modes, the standard operations (e.g., displacement operation, phase rotation, one-mode squeezing, beam splitting, and two-mode squeezing) are all Gaussian operations, which can only transform Gaussian states into Gaussian states \cite{Braunstein2005,Lloyd1999}. However, universal control of a single bosonic mode can be achieved by adding a single nonlinear operation \cite{Lloyd1999}. When such a direct nonlinear operation is difficult to realize directly, it is still possible to implement an indirect nonlinear interaction by coupling the bosonic mode to a finite-level ancilla. Moreover, quantum non-demolition (QND) measurement of the ancilla enables measurement-based feedback control and therefore arbitrary operation on the bosonic mode.
Here, we will review recent advances in the approaches for universal control and arbitrary operation of bosonic modes (Table \ref{Table}), including unitary control, quantum feedback control, driven-dissipative control and holonomic control (Fig. \ref{control}a). In the first two approaches, an ancilla qubit is coupled to a single bosonic mode to introduce nonlinear interaction and feedback control, while in the remaining two approaches, a special coupling between the bosonic mode and reservoir or a special Hamiltonian of the bosonic mode is engineered to support some stabilized manifold, consisting of all coherent superpositions of multiple steady states that are free of any nonunitary effect caused by the reservoir.

\newcommand{\tabincell}[2]{\begin{tabular}{@{}#1@{}}#2\end{tabular}}
\begin{table*}[]
\centering
\caption{Recent theoretical and experimental advances in quantum control of bosonic modes in circuit QED. }
\centering
%\tbl{Recent theoretical and experimental advances in quantum control of bosonic modes in circuit QED.}
\begin{tabular}{p{5.0cm}| p{5.2cm}| p{5.6cm}}
%\begin{tabular}{c|c|c}
\hline
\hline
%\tabincell{c}{\textbf{Unitary control}\\Control via nonlinear ancilla} & \tabincell{c}{\textbf{Unitary \& Feedback}\\Quantum adaptive control} &  \tabincell{c}{\textbf{Unitary \& Dissipation}\\Quantum Zeno dynamics}  &  \tabincell{c}{\textbf{Only dissipation}\\Holonomic quantum control} \\
\tabincell{c}{\textbf{Unitary control}} & \tabincell{c}{\textbf{Unitary \& Feedback}} &  \tabincell{c}{\textbf{Unitary \& Dissipation}} \\

\tabincell{c}{Ancilla-induced nonlinearity} & \tabincell{c}{Quantum adaptive control} &  \tabincell{c}{Quantum Zeno dynamics}\\
\hline
  \tabincell{l}{Theory:\\-~SNAP gate \cite{Krastanov2015}\\-~Optimal control \cite{Khaneja2005,Heeres2017}\\-~E-SWAP gate \cite{Lau2016}\\-~CPHASE gate \cite{Wang2020} \\ \\ \\ }
& \tabincell{l}{Theory:\\-~CPTP maps \cite{Shen2017}\\-~Teleported gate \cite{Gottesman1999}\\-~ET gate \cite{Vy2013,Kapit2018}\\-~PI gate \cite{Ma2019}\\ \\ \\ }
& \tabincell{l}{Theory:\\-~Dissipative cat \cite{Mirrahimi2014}\\-~Kerr cat \cite{Puri2017}\\-~FT syndrome detection \cite{Puri2019}\\ -~Bias-preserving Kerr cat \cite{Puri2019b} \\ -~Bias-preserving dissipative cat \cite{Guillaud2019}\\-~Holonomic gate$^a$ \cite{Albert2016a} \\ }
\\
 \hline
  \tabincell{l}{Experiments:\\-~SNAP gate \cite{Heeres2015}\\-~Optimal control \cite{Heeres2017}\\-~CNOT gate \cite{Rosenblum2018a} \\-~CZ gate \cite{Xu2020}\\-~E-SWAP gate \cite{Gao2019}\\ \\ }
& \tabincell{l}{Experiments:\\-~QEC \cite{Ofek2016,Hu2019,Campagne-Ibarcq2019}\\-~CPTP simulation \cite{Hu2018b,Cai2020}\\-~Teleported CNOT gate \cite{Chou2018}\\-~FT parity measurement \cite{Rosenblum2018b}\\-~PI SNAP gate \cite{Reinhold2019} \\-~ET phase gate \cite{Ma2019b} }
& \tabincell{l}{Experiments:\\-~Dissipative cat \cite{Leghtas2015,Touzard2018,Lescanne2020}\\-~Kerr cat\cite{Grimm2020} \\ \\ \\ \\  \\}   \\
\hline
\hline
\end{tabular}
\\
\footnotesize{$^a$ This scheme uses only dissipation.} \\
\label{Table}
\end{table*}

The physical platform we consider is circuit quantum electrodynamics (circuit QED) \cite{Blais2004,Wallraff2004,Schoelkopf2008,Blais2020,Cai2021,Joshi2020,Blais2021}, which is an analog of cavity QED \cite{Haroche2013} using superconducting circuits \cite{Devoret2013,Haroche2020}. Cavity QED engineers the environment of the atoms by placing them in a cavity that supports only discrete bosonic modes of the electromagnetic field. Examples of cavity QED systems include alkali atoms in optical cavities \cite{Mabuchi2002} and Rydberg atoms in microwave cavities \cite{Nogues1999}. Circuit QED uses superconducting qubits as artificial atoms coupled to microwave resonators. A key advantage of circuit QED is the extremely strong coupling between the superconducting qubits and the cavity.
Corresponding to the two encoding schemes based on qubits or bosonic modes, there are two main routes in superconducting quantum computing with circuit QED.

Qubits can be encoded into the first two levels of superconducting artificial atoms, such as the most widely used transmons \cite{Koch2007,Paik2011,Rigetti2012}), while the cavity resonators are used for qubit readout. Arbitrary single qubit rotations can be realized with resonant voltage drives at the qubit frequencies \cite{Blais2004,Blais2007}, and gate errors can be reduced below $10^{-3}$ by pulse shaping techniques \cite{Barends2014,Chen2016}. Two-qubit gates can be realized by either capacitive coupling \cite{Barends2014} or using the resonator as a mediator \cite{Blais2004,Blais2007,Majer2007}, with current error rates being less than one percent \cite{Sung2020,Samach2021}. Recent experimental developments include implementations of quantum search algorithms \cite{DiCarlo2009}, quantum teleportation \cite{Steffen2013}, simulations of topological transitions \cite{Roushan2014}, digitized adiabatic quantum computing \cite{Barends2016}, variational quantum algorithms \cite{Kandala2017}, supervised learning with quantum-enhanced feature spaces \cite{Havlicek2019}, quantum reservoir engineering \cite{MaR2019}, quantum walks \cite{Gong2021}, advances towards quantum error correction \cite{Kelly2015,Takita2017,Andersen2020,McEwen2021}, building cloud-based devices and demonstrating quantum supremacy with several tens of qubits \cite{Arute2019}.

Alternatively, a storage cavity resonator as a bosonic mode with long coherence time can encode the quantum information, while the transmon qubits can act as ancillas to aid universal control of the storage cavity (Fig. \ref{control}b, c). Such an encoding scheme can make use of various bosonic QEC codes, which are hardware-efficient compared to more standard qubit-based codes such as surface codes. Moreover, bosonic QEC codes often have specific noise resilience \cite{Guillaud2019}, and therefore can be concatenated with conventional QEC codes to reduce the hardware overhead \cite{Guillaud2019,Noh2020,Terhal2020}. Recently there has been significant experimental progress in bosonic QEC. QEC based on cat codes or binomial codes in superconducting cavities have reached or approached the break-even point \cite{Ofek2016,Hu2019}, at which the lifetime of the logical qubit exceeds that of the single best physical qubit within the logical qubit. The encoding based on GKP codes has also been demonstrated in trapped-ion mechanical oscillators \cite{Fluhmann2019,Neeve2020} and superconducting cavities \cite{Campagne-Ibarcq2019}. However, compared to conventional one-qubit and two-qubit control, universal control of single and multiple bosonic modes requires the introduction of nonlinearity and therefore is more complex. This will be the main topic of this review.

This review is organized as follows. In Sec. \ref{unitary}, we review the universal control of a single bosonic mode with the aid of an ancilla qubit dispersively coupled to it. Then we introduce, in Sec. \ref{feedback}, the extension from the universal unitary control to quantum feedback control and arbitrary quantum channel construction for the bosonic mode by QND measurement of the ancilla. In Sec. \ref{dissipative}, it is shown that reservoir engineering and Hamiltonian engineering can be promising strategies to realize universal quantum computation in some unitarily evolving subspace of the bosonic mode. In Sec. \ref{holonomic}, the combination of reservoir engineering and holonomic quantum control is introduced to realize universal control of bosonic modes. Then in Sec. \ref{multimode}, we introduce the quantum control schemes to entangle different bosonic modes for universal quantum computation. In Sec. \ref{summary}, we briefly summarize the review and outline some future directions for quantum control of the bosonic modes. For convenience, we take the reduced Plank constant as $\hbar=1$ throughout this review.

\section{Unitary quantum control} \label{unitary}

Quantum control of a single bosonic mode (typically a harmonic oscillator) can be achieved in the coupled qubit-oscillator system with a qubit as an ancilla. Many theoretical and experimental works were devoted to preparing arbitrary oscillator states assisted by an ancilla qubit with Jaynes-Cummings (JC) coupling \cite{Law1996,Brattke2001,Leibfried2003,Houch2007,Hofheinz2009}. However, it is more challenging to achieve universal control of the oscillator, which usually needs a multi-level ancilla \cite{Santos2005}, slow adiabatic evolutions \cite{Strauch2012} or a large number of control operations \cite{Mischuck2013}.

In circuit QED, the transmon (as an anharmonic oscillator) can act as an ancilla to aid the control of cavity bosonic modes (as harmonic oscillators). If the ancilla and a single oscillator are strongly off-resonant with the detuning much larger than their coupling strength, we arrive at the dispersive Hamiltonian \cite{Schuster2007,Kirchmair2013}
\begin{eqnarray}\label{disp}
   H_0=\sum_{j=0}^{\infty}(\Lambda_j-j\chi a^{\dagger}a)|j\rangle\langle j|+\omega_{\rm C}a^{\dagger}a,
   %H_0=\omega_{\rm T}|e\rangle\langle e|+\omega_{\rm C}a^{\dagger}a -\chi a^{\dagger}a|e\rangle\langle e|,
\end{eqnarray}
where $|j\rangle$ labels the eigenstate of the ancilla, $\Lambda_j$ is the eigenenergy, $\chi$ is the dispersive coupling strength, $\omega_{\rm C}$ is the oscillator frequency, and $a$ ($a^{\dagger}$) is the annihilation (creation) operator of the oscillator excitation. Note that here we have neglected the weak anharmonicity of the cavity modes. The anharmonicity of the ancilla ($\Lambda_j-\Lambda_{j-1}\neq\Lambda_{j+1}-\Lambda_{j}$) makes it possible to selectively drive specific ancilla transitions, so the infinite-dimensional ancilla can often be truncated to a finite-dimensional one. Below the lowest ancilla eigenstates $\{|0\rangle,|1\rangle,|2\rangle\cdots\}$ are denoted as $\{|g\rangle,|e\rangle,|f\rangle\cdots\}$, and the eigenenergy difference between $|e\rangle$ and $|g\rangle$ is denoted as $\omega_{\rm T}=\Lambda_1-\Lambda_0$.

\begin{figure*}
\includegraphics[width=6.5in]{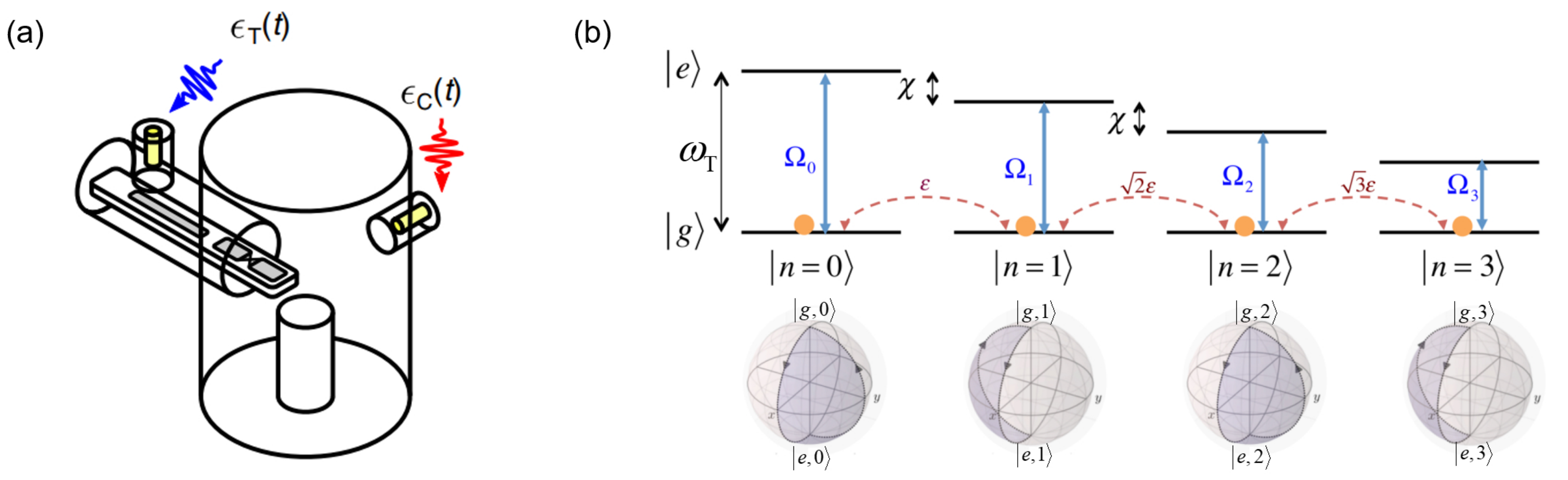}
\includegraphics[width=6.5in]{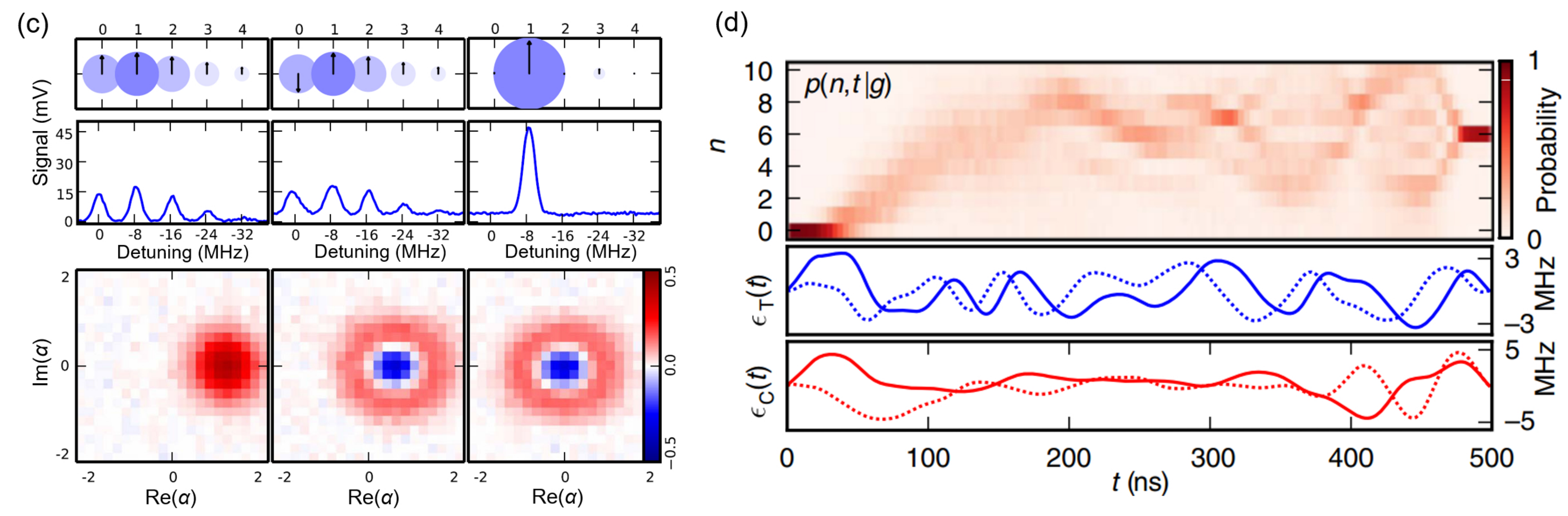}
\caption{ Universal unitary control of a harmonic oscillator via an ancilla qubit. (a) Schematic drawing of the experimental circuit QED system. A $\lambda/4$ coax-stub cavity resonator is coupled to a transmon and readout resonator on a sapphire substrate. Input couplers close to the transmon and cavity deliver the respective time-dependent microwave control fields $\varepsilon_{\rm T}(t)$ and $\varepsilon_{\rm C}(t)$. (b) Schematic of universal control of the qubit-oscillator system via displacements and SNAP gates. A weak displacement operation (red dashed arrows) couples the states $|g,n-1\rangle$ and $|g,n\rangle$ with strength $\sqrt{n}\varepsilon$ for all $n$. The SNAP gate (blue solid arrows) can simultaneously accumulate different geometric phases $\{\varphi_n\}$ to states $\{|g,n\rangle\}$. Here we adopt the rotating frame associated with $\omega_{\rm C} a^{\dagger}a$ so that the states $\{|g,n\rangle\}$ have the same energy. (c) Experimental demonstration of the control strategy in (b) (separately acting on the transmon and the cavity). Phasor representation, transmon spectrum, and Wigner function are shown after each of the steps in the 1-photon Fock state creation experiment. In the phasor representation, the arrow corresponds to the complex amplitude $c_n$ of the initial cavity state $|\psi\rangle=\sum_n c_n|n\rangle$ and the area of the circle is proportional to $|c_n|^2$. The qubit spectrum refers to the ancilla transmon transition frequency dependent on the number of photons in the cavity. (d) Experimental demonstration of control strategy based on numerical GRAPE algorithms (acting on both the transmon and the cavity simultaneously). Lower panel: optimized transmon and oscillator control waveforms of length approximately $2\pi/\chi$ to take the oscillator from vacuum to the 6-photon Fock state. Solid (dotted) lines represent the in-phase (quadrature) field component. Upper panel: oscillator photon-number population trajectory as a function of time conditioned on the transmon in $|g\rangle$. A complex trajectory occupying a wide range of photon numbers is taken to perform the intended operation. Reprinted with permission from \cite{Krastanov2015,Heeres2015,Heeres2017}.}
\label{unitary-fig}
\end{figure*}

The dispersive Hamiltonian can be interpreted from two different perspectives (Fig. \ref{unitary-fig}b). On the one hand, the oscillator frequency has a shift dependent on the ancilla state. This ancilla-state-dependent shift of the cavity leads to changes in the amplitude and phase of photons reflected from or transmitted through the cavity and therefore enables a QND measurement on the ancilla state \cite{Blais2004,Wallraff2004}. On the other hand, the ancilla transition frequency has a shift proportional to the oscillator excitation number. In the strongly dispersive regime of circuit QED, the ancilla frequency shift is much larger than the cavity line width and ancilla line width, and therefore the ancilla spectrum is split into a series of separately resolved peaks, representing the distribution of photon numbers within the driven cavity \cite{Schuster2007}. Moreover, for quantum control of the oscillator, such a strongly-dispersive coupling regime makes it possible to selectively address the ancilla if and only if the oscillator is in a specific number state, hence providing new opportunities for universal control of the oscillator.

Typically we can achieve universal unitary control and quantum measurements of the ancilla, but only limited unitary control on the oscillator, so the key point is to use the ancilla to realize some other unitary control on the oscillator to achieve universal control. Below we introduce two schemes: the unitary control either separately acts on the ancilla or the oscillator and then is combined, or acts on the both of them simultaneously (Fig. \ref{unitary-fig}a).

\subsection{Displacement operations and SNAP gates}\label{sec-SNAP}

The first scheme for universal control of the oscillator is to separately apply unitary control on the subsystems (either the ancilla or the oscillator) and then combine them \cite{Krastanov2015,Heeres2015}. The unitary control on the ancilla may indirectly realize some unitary operations on the oscillator if we make appropriate pre-section and post-selection of the ancilla state. Then combining these indirect operations with the direct ones, we may realize universal control of the oscillator.

One common kind of direct unitary transformation on the oscillator is the displacement operation
\begin{eqnarray}
    D(\alpha)=\exp(\alpha a^{\dagger}-\alpha^{*}a),
\end{eqnarray}
which can be generated by a linear drive on the cavity $H_{\rm C}=\epsilon_{\rm C}(t)e^{i\omega_{\rm C} t}a^{\dagger}+\rm{H.c.}$ with $\alpha=-i\int \epsilon_{\rm C}(t)dt$ . However, the displacement operation alone is not universal, i.e. it cannot generate arbitrary operations on the oscillator. To see this, note that the displacement operation can only prepare a coherent state from the vacuum, i.e. $|\alpha\rangle=D(\alpha)|0\rangle=\exp(-|\alpha|^2/2)\sum_{n=0}^{\infty}(\alpha^n/\sqrt{n!})|n\rangle$, while universal control requires that any given target state can be prepared from any initial state including the vacuum state.

The dispersive coupling of the oscillator to the ancilla introduces a nonlinear term for the oscillator Hamiltonian, and make it possible to realize indirect control on the oscillator, such as the selective number-dependent arbitrary phase (SNAP) gate,
\begin{eqnarray}
S\left( {\vec \varphi } \right) =\sum_{n=0}^\infty e^{i\varphi_n}|n\rangle\langle n|,
%= \prod\limits_{n = 0}^\infty  {{e^{i{\varphi _n}\left| n \right\rangle \left\langle n \right|}}} }.
\end{eqnarray}
which imparts arbitrary phases $\vec \varphi=\{\varphi_n\}_{n=0}^{\infty}$ to the different Fock states of the oscillator. The original proposal to realize SNAP gates is to weakly drive the ancilla with multiple frequency components, $H_{\rm T}=\epsilon_{\rm T}(t) e^{i\omega_{\rm T}t}|g\rangle\langle e|+\text{H.c.}$ with $\epsilon_{\rm T}(t)=\sum_n \Omega e^{i(\phi_n(t)-n\chi t)}$ and $\phi_n(t)$ being time-dependent. If $\Omega\ll \chi$, the driving component with frequency $\omega_q-n\chi$ induces a unitary evolution in the ancilla subspace $\{|g,n\rangle, |e,n\rangle\}=\{|g\rangle,|e\rangle\}\otimes|n\rangle$ with a negligible effect on the rest of the system, while the driving phases \{$\phi_n(t)$\} depending on the oscillator excitation numbers can induce different evolution paths in different ancilla subspaces, as shown in Fig. \ref{unitary-fig}b. Suppose the initial state of the whole system is a product state, $|\psi(0)=|g\rangle\otimes\sum_{n=0}^{N}c_n|n\rangle=\sum_{n=0}^{N}c_n|g,n\rangle$ with $N$ being the truncated oscillator excitation number, then we may let the ancilla undergo cyclic evolutions in each subspace $\{|g,n\rangle, |e,n\rangle\}$ and return to $|g,n\rangle$ at time $\tau$. We can tune $\phi_n(t)$ so that the final state accumulates different geometric phases $\varphi_n$ for different $n$ \cite{Aharonov1987}, i.e. $|\psi(\tau)\rangle=\sum_{n=0}^{N}c_n e^{i\varphi_n}|g,n\rangle$.
For example, we may set $\phi_n(t)=0$ for $t\in[0,\tau/2)$ and $\phi_n(t)=\varphi_n$ for $t\in[\tau/2,\tau]$ with $\tau=\pi/\Omega$ being the Rabi period, and the unitary propagator on the whole system [in the interaction picture associated with the dispersive Hamiltonian in Eq. (\ref{disp})] is
\begin{eqnarray}\label{U-SNAP}
    U(\tau,0)=|g\rangle\langle g|\otimes S(\vec \varphi)+|e\rangle\langle e|\otimes S(-\vec \varphi),
\end{eqnarray}
which implies that the unitary gate on the oscillator is $S(\vec \varphi)$ [$S(-\vec \varphi)$] if the initial ancilla state is $|g\rangle$ ($|e\rangle$).

The original SNAP gate based on the geometric phases can be simplified by first decomposing the above propagator [Eq.(\ref{U-SNAP})] as $U(\tau,0)=U(\tau,\tau/2)U(\tau/2,0)$ with
\begin{subequations}\label{}
\begin{align}
    & U(\tau/2,0)=(|g\rangle\langle e|+|e\rangle\langle g|)\otimes \mathbb{I},\\
    & U(\tau,\tau/2)=|g\rangle\langle e|\otimes S(\vec \varphi)+|e\rangle\langle g|\otimes S(-\vec \varphi),
\end{align}
\end{subequations}
where $\mathbb{I}$ is the identity operator for the oscillator. Note that the first half evolution $U(\tau/2,0)$ causes a flip of the ancilla state while leaving the oscillator state unchanged, and the second half evolution $U(\tau,\tau/2)$ causes a further flip of the ancilla state and produces the SNAP gate on the oscillator at the same time (Fig. \ref{unitary-fig}b). So we may simplify the SNAP gate by applying only the drive during the second half period, which we may call the \textit{simplified SNAP gate}. Moreover, if the simplified SNAP gate is not completed, we have
\begin{align}\label{Usnap}
    &U(\tau/2+\Delta t,\tau/2) \nonumber \\
    &=\cos\theta(|g\rangle\langle g|+|e\rangle\langle e|)\otimes \mathbb{I} \nonumber \\
    &-i\sin\theta\left[|g\rangle\langle e|\otimes S(\vec \varphi)+|e\rangle\langle g|\otimes S(-\vec \varphi)\right],
\end{align}
where $\Delta t\in[0,\tau/2]$ and $\theta=\Omega\Delta t$. In this case, the pre-selection and post-selection of the ancilla state induces either the identity operation or SNAP gate on the oscillator. For example, $P_eU(\tau/2+\Delta t,\tau/2)P_e=\mathbb{I}$ and $P_gU(\tau/2+\Delta t,\tau/2)P_e=S(\vec \varphi)$ with $P_m=|m\rangle\langle m|$ ($m=g,e$).
Note that in the above discussions, we consider the limiting case $\Omega/\chi\rightarrow 0$, while in practice $\Omega/\chi$ is finite and causes deviations from the ideal SNAP gates \cite{Krastanov2015,Heeres2015}. Nevertheless, it is possible to minimize such gate errors due to finite $\Omega/\chi$ by optimizing the detunings and pulse shapes of the multi-frequency drive on the ancilla \cite{Wang2020}. Besides the resonant driving approach for SNAP gates, a photon-number dependent Hamiltonian of the oscillator can also be engineered by off-resonantly driving the ancilla with multiple frequencies \cite{Wang2020}.

It has been demonstrated that universal control of an oscillator can be achieved by combining the displacement operations $D(\alpha)$ and the SNAP gates $S(\vec \varphi)$ \cite{Krastanov2015}, since the generators of $D(\alpha)$ and $S(\vec \varphi)$ and the commutators between these generators generate the full Lie algebra $\mathfrak{u}(N)$ for any truncated oscillator space $\{|0\rangle, \cdots, |N-1\rangle\}$ \cite{Lloyd1999}. As an example, we show in Fig. \ref{unitary-fig}c that a Fock state $|1\rangle$ of the oscillator can be created by applying the operation $D(\beta_2)S(\vec \varphi)D(\beta_1)$, where $\vec \varphi$ is fixed to be $(\pi,0,0,\cdots)$ while the displacement parameters $\beta_1$, $\beta_2$ are obtained by numerical optimization. A systematic method was presented in \cite{Krastanov2015} to construct an arbitrary unitary operation in any truncated oscillator space. With this method, the number of operations to prepare the oscillator Fock state $|n\rangle$ can be significantly decreased from $\mathcal{O}(n)$ to $\mathcal{O}(\sqrt{n})$. Recently a more efficient scheme by parameter optimization has been proposed to implement a broad range of cavity control with only 3 to 4 SNAP gates \cite{Fosel2020}. Nevertheless, it is still an open problem to find the optimal way for decomposing an arbitrary target unitary into displacement operations and SNAP gates.

\subsection{Universal control by numerical optimization algorithms}
The previous analytic approach based on displacement operations and SNAP gates implicitly assumes that the cavity drive [$\epsilon_{\rm C}(t)$] and transmon drive [$\epsilon_{\rm T}(t)$] are never applied simultaneously, which makes the evolution more tractable. However, to find more efficient control schemes, it is better to include the possibility of simultaneously driving both the ancilla and the oscillator. The arbitrary control field can take the form
\begin{eqnarray}\label{}
    H_{\rm CT}=\epsilon_{\rm C}(t)e^{i\omega_{\rm C} t}a+\epsilon_{\rm T}(t)e^{i\omega_{\rm T} t}|g\rangle\langle e|+\rm{H.c.},
\end{eqnarray}
where $\epsilon_{\rm C}(t)$, $\epsilon_{\rm T}(t)$ are arbitrary complex-valued functions of time. The exact form of the control field can be obtained by numerical optimization algorithms \cite{Heeres2017}, such as the Gradient Ascent Pulse Engineering (GRAPE) method \cite{Khaneja2005,deFouquieres2011}. The basic procedure of the GRAPE method is as follows: (1) specify the target unitary $U$ and the evolution time $\tau$; (2) discretize the total time $\tau$ into $M$ equal steps of duration $\Delta t=\tau/M$, and during each step the control amplitudes are constant; (3) make an initial guess of the control amplitudes, then calculate the fidelity between the implemented unitary and target unitary, and also the gradient of the fidelity with respect to each variation of the control amplitude in each time step; (4) adapt the control amplitudes according to the fidelity gradient, and repeat step (3) until a local maximum of the gate fidelity is achieved.

Both $\epsilon_{\rm C}(t)$ and $\epsilon_{\rm T}(t)$ can be optimized with GRAPE to achieve universal control of the cavity. The numerical optimized pulses thus obtained have been extensively used in experiments to control superconducting cavity modes \cite{Ofek2016,Axline2018,Rosenblum2018a,Hu2019}. As an example, we show in Fig. \ref{unitary-fig}d the control amplitudes $\epsilon_{\rm C}(t)$, $\epsilon_{\rm T}(t)$ to prepare the cavity state from the vacuum state $|0\rangle$ to the Fock state $|6\rangle$. With this approach, Heeres \textit{et al.} \cite{Heeres2017} have also realized a universal set of gates on the logical qubit based on error-correcting cat codes in a cavity. Compared with the SNAP gate that takes a rather long time $2\pi/\Omega$ due to $\Omega/\chi\ll1$, the logical gates based on GRAPE algorithm take a much shorter time $2\pi/\chi$.

\subsection{Other approaches}
Besides the above schemes, there are various other approaches to control harmonic oscillators via ancilla-induced nonlinearity. One approach is called photon blockade control, in which the frequency of $\epsilon_{\rm T}(t)$ is set as $\omega_{\rm T}-N\chi$ to drive resonantly the transition $|g,N\rangle\leftrightarrow |e,N\rangle$, therefore blockading the population transfer between the cavity subspace $\{|0\rangle, |1\rangle, \cdots, |N-1\rangle\}$ and the rest of the cavity Hilbert space \cite{Bretheau2015}. Then universal control of the $N$-level qudit can be realized by optimizing $\epsilon_{\rm C}(t)$ with GRAPE \cite{Schirmer2001}, which has been experimentally demonstrated in \cite{Chakram2020}. In another approach, using a single transmon as the central processor, universal quantum operations have been realized between arbitrary eigenmodes of a linear array of coupled superconducting resonators, realizing a random access quantum information processor \cite{Naik2017}.

It should be mentioned that the weak point of the transmon as an ancilla is its small dispersive coupling strength with the cavity modes. This limits the control fidelity of both SNAP gates and blockade control. Such limitations may be overcome in the future by using better ancillas. For example, with an ancilla oscillator, a superconducting oscillator can have stimulated nonlinearity by a three-wave interaction, enabling control of the single-photon manifold at rates faster than the dispersive protocols \cite{Vrajitoarea2020}. Other possible better ancillas include the C-shunt flux qubit with large anharmonicity \cite{Abdurakhimov2019} and the fluxonium with millisecond coherence time \cite{Manucharyan2009,Somoroff2021}.

\section{Quantum feedback control} \label{feedback}
In the last section, the system (an oscillator and an ancilla) as a whole are assumed to be a closed system and therefore can be sufficiently described by unitary dynamics. However, the inevitable coupling of the system to the environment typically induces non-unitary evolutions of the system, which can be fully characterized by completely positive and trace preserving (CPTP) maps \cite{Nielsen2000,Wilde2013} (also called quantum operations or quantum channels). Hence, it is important to systematically extend the quantum control techniques from a closed system to an open quantum system. In this section, we will show that an arbitrary CPTP map of the system can be constructed by coupling the system to an ancilla qubit with QND readout and quantum feedback control.

Feedback control, where information about the system state is fed back to the controller for correction, is widely used in classical control theory. However, its extension to the quantum world is nontrivial \cite{Wiseman2010}, since a quantum measurement of the system will inevitably affect the quantum state of the system. Quantum feedback control generally falls into two categories: measurement-based feedback control \cite{Wiseman1993} and coherent feedback control \cite{Lloyd2000}. Below we will show that the measurement-based approach can be used to construct arbitrary CPTP maps and realize robust quantum operations.

\begin{figure*}
\includegraphics[width=6.5in]{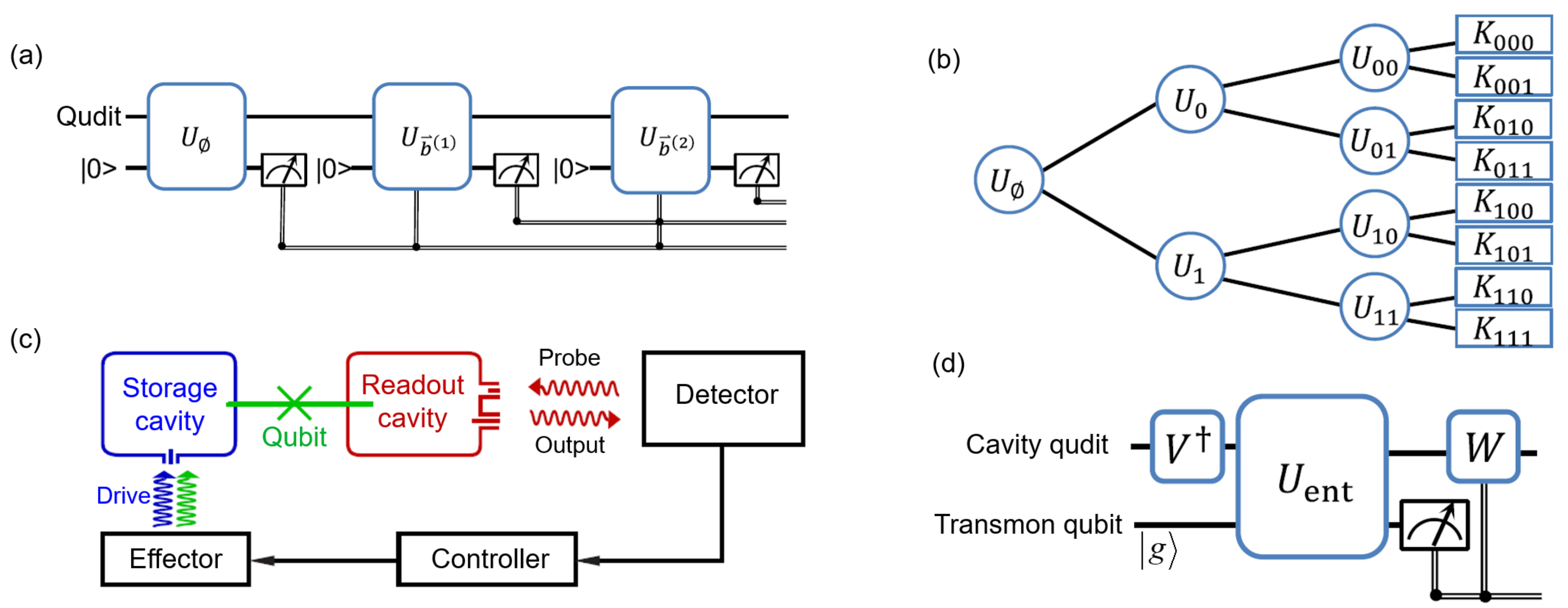}
\caption{ Arbitrary CPTP map construction with quantum feedback control. (a) Binary tree representation with depth $L = 3$. The Kraus operators $K_{b^{(L)}}$ are associated with the leaves of the binary tree, $b^{(L)}\in\{0,1\}^L$. The system-ancilla joint unitary to apply in $l$th round $U_{b^{(l)}}$ depends on the previous ancilla readout record $b^{(l)} = (b_1 b_2\cdots b_l )\in\{0,1\}^l$ associated with a node of the binary tree. (b) Schematic setup of a circuit QED system used for constructing an arbitrary quantum channel. (c) Quantum circuit for arbitrary channel construction. The dimension of the system $d$ can be arbitrary and the circuit depth depends only on the Kraus rank of the target channel. (d) The quantum circuit to implement an arbitrary Kraus rank-2 channel with the circuit QED system. Reprinted with permission from \cite{Shen2017}.}
\label{CPTP}
\end{figure*}

\subsection{Arbitrary CPTP map construction}
A CPTP map can be described by the Kraus representation \cite{Wilde2013}
\begin{eqnarray}\label{}
    \mathcal{\varepsilon}(\rho)=\sum_{i=1}^N K_i \rho K_i^{\dagger},
\end{eqnarray}
where $\rho$ is the density matrix of the system we consider and $\{K_i\}_{i=1}^{N}$ is the set of Kraus operators satisfying $\sum_{i=1}^N K_i^{\dagger}K_i=\mathbb{I}$ to preserve the trace of $\rho$. The Kraus representation is not unique, since a new set of Kraus operators $\{F_i\}_{i=1}^{N}$ can be constructed with any $N\times N$ unitary matrix $U$, $F_i=\sum_jU_{ij}K_j$, characterizing the same CPTP map. The minimum number
of Kraus operators is called the \textit{Kraus rank} of the CPTP map, and is no larger than $d^2$ with $d$ being the Hilbert space dimension of the system.

\subsubsection{Construction of CPTP maps with arbitrary Kraus rank}

For the construction of arbitrary CPTP maps, Lloyd and Viola first showed that it is sufficient to repeatedly apply Kraus rank-2 channels in an adaptive fashion \cite{Lloyd2001}, but they did not consider efficient construction with a low-depth quantum circuit. Recently Shen \textit{et al}. have extended the binary-tree construction for arbitrary positive operator-valued measure (POVM) \cite{Andersson2008} to an efficient protocol for CPTP map construction \cite{Shen2017}. In this protocol, a CPTP map with Kraus rank-$N$ can be constructed with an ancilla qubit by the lowest possible circuit depth $L=\,\log_2 N\!$, where each round of operation consists of one joint unitary of system and ancilla and one QND measurement on the ancilla qubit. Below we will briefly introduce such a binary-tree construction for CPTP maps.

Let us first consider the construction of a rank-2 CPTP map with Kraus operators $\{K_0,K_1\}$, which can be achieved by only one round of operation: (1) initialize the ancilla qubit in $|0\rangle$ (the qubit state basis being $\{|0\rangle,|1\rangle\}$); (2) perform a joint unitary operation $U\in SU(2d)$ with $d$ being the dimension of the system; (3) discard or trace over the ancilla qubit. The key point is to design $U$ so that its $d\times d$ submatrices satisfy $\langle 0|U|0\rangle=K_0$, $\langle 1|U|0\rangle=K_1$.

The quantum circuit to implement a rank-$N$ CPTP map with Kraus operators $\{K_0,\cdots, K_N\}$ consists of $L=\,\log_2 N\!$ rounds of operations (Fig. \ref{CPTP}a). Each round of operation includes: (1) initialization of the ancilla qubit in $|0\rangle$; (2) joint unitary gate over the system and ancilla (conditional on the measurement outcomes from previous rounds), (3) QND readout of the ancilla, and (4) storage of the classical measurement outcome for later use. The $l$th round unitary gate $U_{b^{(l)}}$ is represented by the node of the binary tree $b^{(l)} = (b_1 b_2\cdots b_l )\in\{0,1\}^l$ with $l=0,\cdots,L-1$, while the Kraus operators $K_{b^{(L)}}$ are associated with the leaves of the binary tree $b^{(L)}\in\{0,1\}^L$ (Fig. \ref{CPTP}b). A systematic way is presented in \cite{Shen2017} to design the nodes $U_{b^{(l)}}$ so that the leaves of the binary tree are exactly the desired Kraus operators, $K_{b^{(L)}}=K_i$ for $i=(b_1 b_2\cdots b_L)_2+1\leq N$ and $K_{i>N}=0$ [$(\cdots)_2$ denotes a binary number]. Arbitrary quantum channels can also be constructed in the quantum circuit model including controlled-not (CNOT), single-qubit gates and partial
trace operations on the qubits and any ancilla, and with free single-qubit gates the minimum number of CNOT gates has been found in \cite{Iten2017}.

\subsubsection{Physical implementation with circuit QED}

Circuit QED in the dispersive regime is a promising platform to implement the arbitrary CPTP map construction. The transmon qubit acts as the ancilla (the transmon state $|g/e\rangle$ corresponds to the ancilla state $|0/1\rangle$ in the last subsection), and a $d$-dimensional subspace (e.g., the lowest $d$ Fock states) of the storage cavity with high-quality-factor (high-$Q$) acts as the qudit. The QND readout of the transmon qubit can be realized by coupling a readout cavity with low-$Q$ to the transmon. Then the readout result is fed back to a controller that induces an effector to implement the conditional control on the qudit (Fig. \ref{CPTP}c).

Similar to the SNAP gates, we can implement the following entangling unitary gate for the whole system including the transmon and the cavity,
\begin{eqnarray}\label{}
    U_{\rm ent}(\vec{\theta})=\prod_{n=0}^d \exp(-iY_n\theta_n/2),
\end{eqnarray}
where $\vec{\theta}=(\theta_0,\cdots,\theta_d)$, and $Y_n=-i|g,n\rangle\langle e,n|+{\rm H.c.}$ is the the Pauli-$Y$ operator for the two-dimensional subspace $\{|g,n\rangle, |e,n\rangle\}$. The drive on the transmon for the above gate is $H_{\rm ent}=\sum_n \Omega_n e^{-i(\omega_{\rm T}-n\chi)t}|g\rangle\langle e|+\text{H.c.}$, where the driving amplitude $\Omega_n$ and the gate time $\tau$ should satisfy $\theta_n=2\Omega_n\tau$. This entangling gate produces a CPTP map with Kraus operators $\{S_g, S_e\}$ with $S_g={\rm diag}(\cos\theta_1,\cdots,\cos\theta_d)$ and $S_e={\rm diag}(\sin\theta_1,\cdots,\sin\theta_d)$. If we precede $U_{\rm ent}$ with a unitary $V^{\dagger}$ acting on the qudit alone and perform an conditional unitary $W=|g\rangle\langle g|\otimes W_g+|e\rangle\langle e|\otimes W_e$ after $U_{\rm ent}$, the entangling gate becomes $U_{\rm ent}'=WU_{\rm ent}V^{\dagger}$ (Fig. \ref{CPTP}d), which is known as the ``cosine-sine" decomposition \cite{Shende2006} that can decompose an arbitrary unitary into CNOT and single-qubit gates. The Kraus operators corresponding to $U_{\rm ent}'$ are $\langle g|U_{\rm ent}'|g\rangle=W_g S_g V^{\dagger}$, $\langle e|U_{\rm ent}'|g\rangle=W_e S_e V^{\dagger}$, which are singular value decomposition of any operator for the qubit \cite{Nielsen2000} and therefore can simulate any rank-2 CPTP map. Likewise we can use such entangling gates to simulate the CPTP map with any Kraus rank.

Recently there have been several experiments for quantum channel simulations in various platforms, including trapped ions \cite{Fluhmann2019}, nuclear mangnetic resonance (NMR) system \cite{Xin2017} and IBM's cloud computer \cite{Wei2018}. In particular, using a scheme similar to the above one, Hu \textit{et al}. \cite{Hu2018b} first realized arbitrary quantum channel simulation for a single photonic qubit in circuit QED. Although this experiment only simulates quantum channels with Kraus rank-2 for a 2-level qubit with one round of adaptive control, a recent experiment has extended the capability to simulate arbitrary rank-16 channels for a $4$-level qudit with 4 rounds of adaptive control \cite{Cai2020}. For the platforms other than circuit QED, the real-time adaptive control is often the main limitation: for trapped ions, it is quite challenging to avoid the recoil problem when performing adaptive measurement for trapped ions; for NMR systems, single-shot readout is not available, so many ancillas must be used to simulate the adaptiveness; the IBM's cloud computer does not allow real-time adaptive control.

The ability to construct an arbitrary CPTP map may have various applications, such as QEC and quantum state initialization/stabilization. For example, the simulated quantum channel enabling QEC can help achieve the Heisenberg limit in quantum metrology \cite{Reiter2017,Zhou2018}, and dissipative quantum circuits consisting of sequences of quantum channels subject to specific constraints can lead to finite-time robust state stabilization \cite{Johnson2017}.

\subsection{Robust quantum operations with adaptive control}

\begin{figure*}
\includegraphics[width=6.5in]{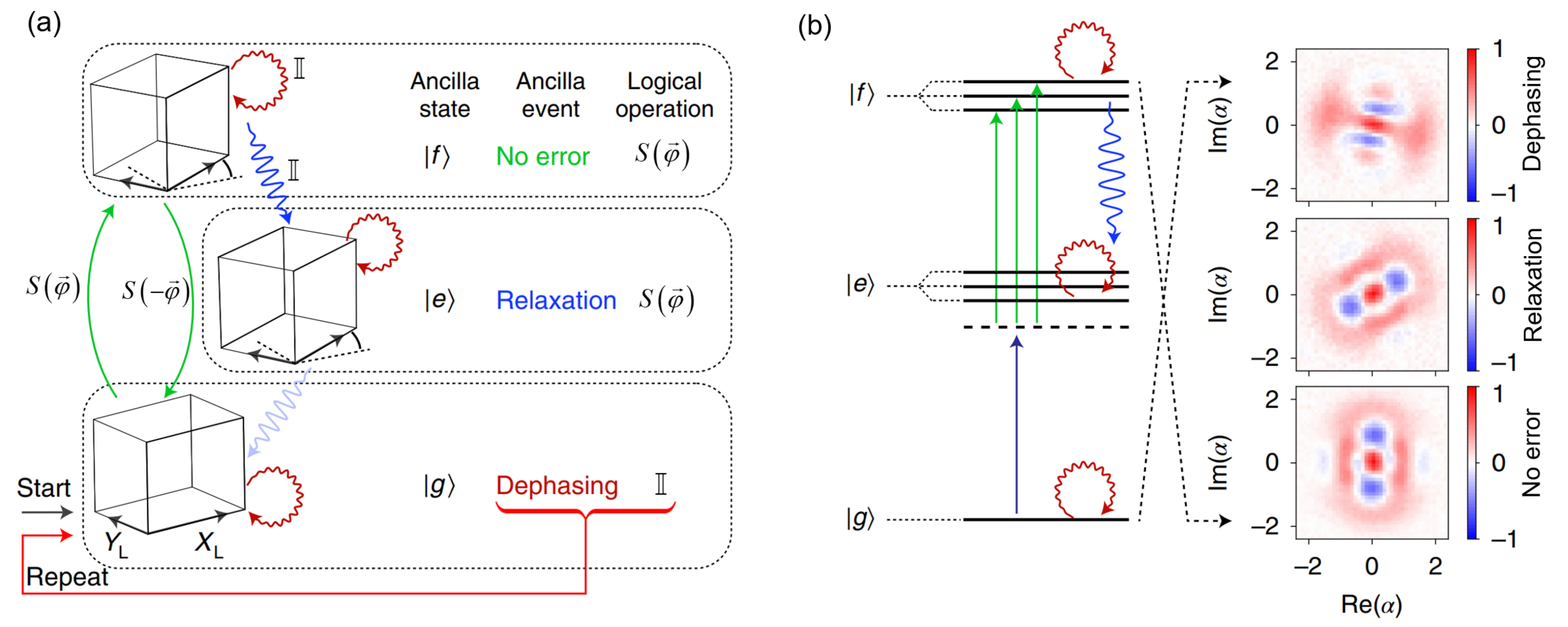}
\includegraphics[width=6.5in]{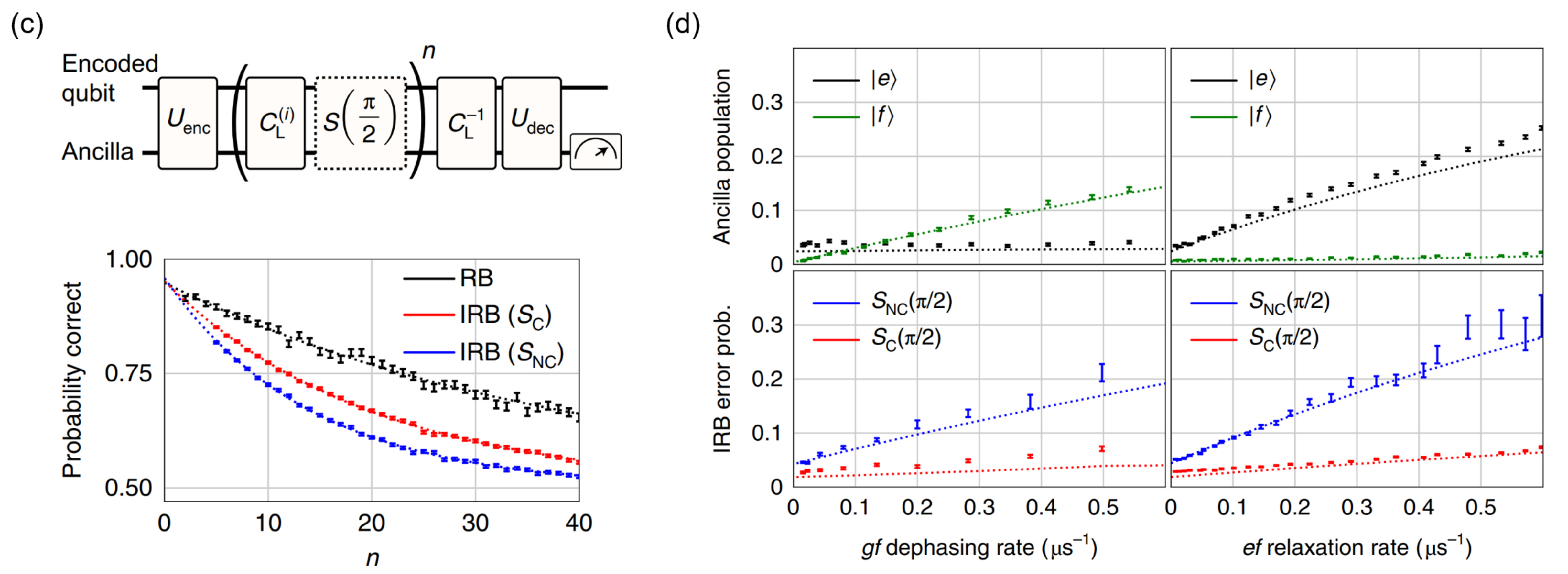}
\caption{ Error-corrected (PI) SNAP gate in circuit QED. (a) PI principle of error-corrected SNAP gate demonstrated by the ancilla transition graph. The ancilla transition from the ground state $|g\rangle$ to the second excited state $|f\rangle$ (green arrows) implements the SNAP gate $S(\vec\varphi)$ on the logical system (boxes), while the reverse ancilla transition from $|f\rangle$ to $|g\rangle$ implements the inverse SNAP gate $S(-\vec\varphi)$. The ancilla transition from any state to itself (red closed loops) produces the identity operation on the logical system. Without ancilla errors, all closed loops in the ancilla transition graph produce the identity operation on the logical system, satisfying the PI condition. With the $\chi$-matching condition in Eq. (\ref{dis-3}), the dominant ancilla relaxation error from $|f\rangle$ to $|e\rangle$ (blue arrows) produces an identity operation on the logical system, still ensuring the PI condition. The ancilla relaxation from $|e\rangle$ to $|g\rangle$ breaks the PI condition but is a second-order error. (b) The SNAP operation for implementing a logical rotation $S(\varphi)=e^{-iZ_L\varphi}$ with $Z_L=|0_L\rangle\langle 0_L|-|1_L\rangle\langle 1_L|$ for the binomial code $\{||0_L\rangle=(|0\rangle+|4\rangle)/\sqrt{2}, |1_L\rangle=|2\rangle\}$. The control consists of applying a Raman drive detuned from the $|g\rangle\leftrightarrow|e\rangle$ transition (blue arrow) as well as a comb of control drives (green arrows), detuned in the opposite sense from the $|e\rangle\leftrightarrow|f\rangle$ transition and separated in frequency by twice the ancilla-cavity dispersive shift $2\chi$. The measured Wigner tomograms of the cavity state, postselected on the final ancilla state following a $|g\rangle\leftrightarrow|f\rangle$ swap (dashed arrows), are shown to the right. (c) Error-corrected SNAP gate performance from randomized benchmarking (RB) and interleaved randomized benchmarking (IRB). The effective gate error probability can be learnt by fitting both the RB and IRB results to an exponential model (dotted lines). The error probability without interleaved logical gates is $\gamma_{\rm RB}$ = 2.5\%$\pm$0.1\% (black), while the error probability associated with the error-corrected S$_{\rm C}$ (non-error-corrected S$_{\rm NC}$) operation is $\gamma_{\rm IRB}-\gamma_{\rm IRB}$ = 2.4\%$\pm$0.1\% (4.6\%$\pm$0.1\%) from the red (blue) curve. (d) Robustness of the error-corrected SNAP gate with added ancilla dephasing ($|f\rangle\langle f|-|g\rangle\langle g|$) and relaxation noise rates ($|e\rangle\langle f|$). In both cases, S$_{\rm C}$ (red markers) is significantly less likely than S$_{\rm NC}$ (blue markers) to translate ancilla errors induced by the added noise into logical errors. The dotted lines are derived from a full quantum simulation using independently measured system parameters. Reprinted with permission from \cite{Reinhold2019}.}
\label{PI-snap}
\end{figure*}

The measurement-based adaptive control can also help achieve robust quantum operations, such as FT quantum measurements and FT quantum gates. Below we show the recent theoretical and experimental advances of FT operations enabled by adaptive control in circuit QED.

In Sec. \ref{unitary}, we have shown that universal control of a bosonic mode can be achieved with the aid of an ideal ancilla. In addition, the ancilla can measure the even-odd excitation number parity of the bosonic mode (as an error syndrome detecting single excitation loss). This can be done by first preparing the ancilla in state $(|g\rangle+|e\rangle)/\sqrt{2}$, evolving with the dispersive Hamiltonian [Eq. (\ref{disp})] for a time $\pi/\chi$, and finally performing Ramsey interferometry on the ancilla to determine its phase. However, ancilla systems are typically more vulnerable to environmental noise, e.g. the transmon coherence time ($\sim\mu$s) is much shorter than the cavity mode coherence time ($\sim$ms), so the ancilla errors (e.g. relaxation error $|g\rangle\langle e|$ and dephasing error $|e\rangle\langle e|-|g\rangle\langle g|$) during the operation time can propagate to the oscillator and corrupt the encoded information irreversibly. This drawback calls for new operation schemes that are FT to these ancilla errors.

A recent experiment shows that the parity measurement of a cavity mode in circuit QED can be made FT to the ancilla transmon errors by using three transmon levels and adaptive control \cite{Rosenblum2018b}. The three-level transmon ($|g\rangle, |e\rangle, |f\rangle$) is coupled to a cavity mode with
\begin{eqnarray}\label{dis-3}
    H_{\rm dis}=-\chi a^{\dagger}a(|e\rangle\langle e|+|f\rangle\langle f|).
\end{eqnarray}
Note that the dispersive coupling strength is the same for the transmon in $|e\rangle$ or $|f\rangle$ ($\chi$-matching condition), which can be realized with an engineered side-band drive \cite{Rosenblum2018b}. The dispersive Hamiltonian commutes with the dominant ancilla relaxation error ($|e\rangle\langle f|$) and also any ancilla dephasing error ($c_g|g\rangle\langle g+c_e|e\rangle\langle e|+c_f|f\rangle\langle f|$ with $c_g,c_e,c_f\in\mathbb{C}$). Such an ancilla error during the measurement is equivalent to an ancilla error at the end, so although the measurement fails if the error happens, the cavity logical state is still well protected and the measurement errors can be overcome by majority voting.

With the same three-level ancilla satisfying the $\chi$-matching condition, the SNAP gates in Sec. \ref{sec-SNAP} can be made FT to the dominant ancilla relaxation error and any dephasing error by adaptive control \cite{Ma2019,Reinhold2019}. Such a SNAP gate is implemented by applying the Hamiltonian that drives the $|g\rangle\leftrightarrow|f\rangle$ transition instead of the $|g\rangle\leftrightarrow|e\rangle$ transition, with the effective Hamiltonian in the interaction picture as
\begin{align}\label{Usnap}
    H_{\rm int}=\Omega\left[|g\rangle\langle f|\otimes S(-\vec \varphi)+|f\rangle\langle g|\otimes S(\vec \varphi)\right].
\end{align}
Without any ancilla error, the logical gate on the cavity with the ancilla going from $|g\rangle$ to $|f\rangle$ is the ideal SNAP gate $S(\vec \varphi)$. With a single ancilla relaxation error $|e\rangle\langle f|$ during the control, the ancilla ends in $|e\rangle$ and the final logical operation is still $S(\vec \varphi)$. With a single ancilla dephasing error (e.g. $|f\rangle\langle f|-|g\rangle\langle g|$) and a projective measurement of the ancilla after the gate, the ancilla may end in $|f\rangle$ with the logical gate still being $S(\vec \varphi)$, or end in $|g\rangle$ with the logical gate being the identity operation (Fig. \ref{PI-snap}a, b). Thus the control protocol can be repeated if the ancilla is measured in $|g\rangle$ until the SNAP gate succeeds. Such error-corrected SNAP gates have recently been experimentally realized \cite{Reinhold2019} with a reduction of the logical gate error by a factor of two in the presence of naturally occurring decoherence, a sixfold suppression of the gate error with increased transmon relaxation rates and a fourfold suppression with increased transmon dephasing rates (Fig. \ref{PI-snap}c, d).

Recent theory shows that the error-corrected SNAP gate belongs to a general class of FT gates on a logical system protected against Markovian ancilla errors, called path-independent (PI) quantum gates \cite{Ma2019}. The PI principle requires that for given initial and final ancilla states, the logical system undergoes a unitary gate independent of the specific ancilla path induced by control drives and ancilla error events. With a certain initial ancilla state, the desired quantum gate on the logical system is successfully implemented for some final ancilla states, while the other final ancilla states herald a failure of the attempted operation, but the logical system still undergoes a deterministic unitary evolution without loss of coherence. So the PI gate on the central system can be repeated until it succeeds. A special class of the PI gates is the error-transparent (ET) gates for a QEC code, theoretically proposed in \cite{Vy2013,Kapit2018} and experimentally demonstrated \cite{Ma2019b} against a specific system error.

The FT measurement and PI gates belong to an interesting class of CPTP maps, called quantum instruments \cite{Shen2017}. For quantum instruments, both the classical measurement outcomes and the post-measurement states of the quantum system are tracked, with the corresponding CPTP map
\begin{align}\label{}
    \mathcal{\varepsilon}_{\rm QI}(\rho)=\sum_{\mu=1}^M \mathcal{\varepsilon}_{\mu}(\rho)\otimes |\mu\rangle\langle \mu|,
\end{align}
where $\{|\mu\rangle\langle \mu|\}_{\mu=1}^{M}$ is a set of $M$ orthogonal projections of the measurement device, and $\{\mathcal{\varepsilon}_{\mu}\}_{\mu=1}^{M}$ is a set of completely positive and trace nonincreasing maps while $\sum_{\mu=1}^{M}\mathcal{\varepsilon}_{\mu}(\rho)$ preserves the trace. For the FT parity measurement, $\{\mathcal{\varepsilon}_{\mu}\}$ contains either the parity measurement channels or the identity channel, while for PI gates $\{\mathcal{\varepsilon}_{\mu}\}$ is a set of unitary channels.

\section{Driven-dissipative control} \label{dissipative}

\begin{figure*}
\includegraphics[width=7in]{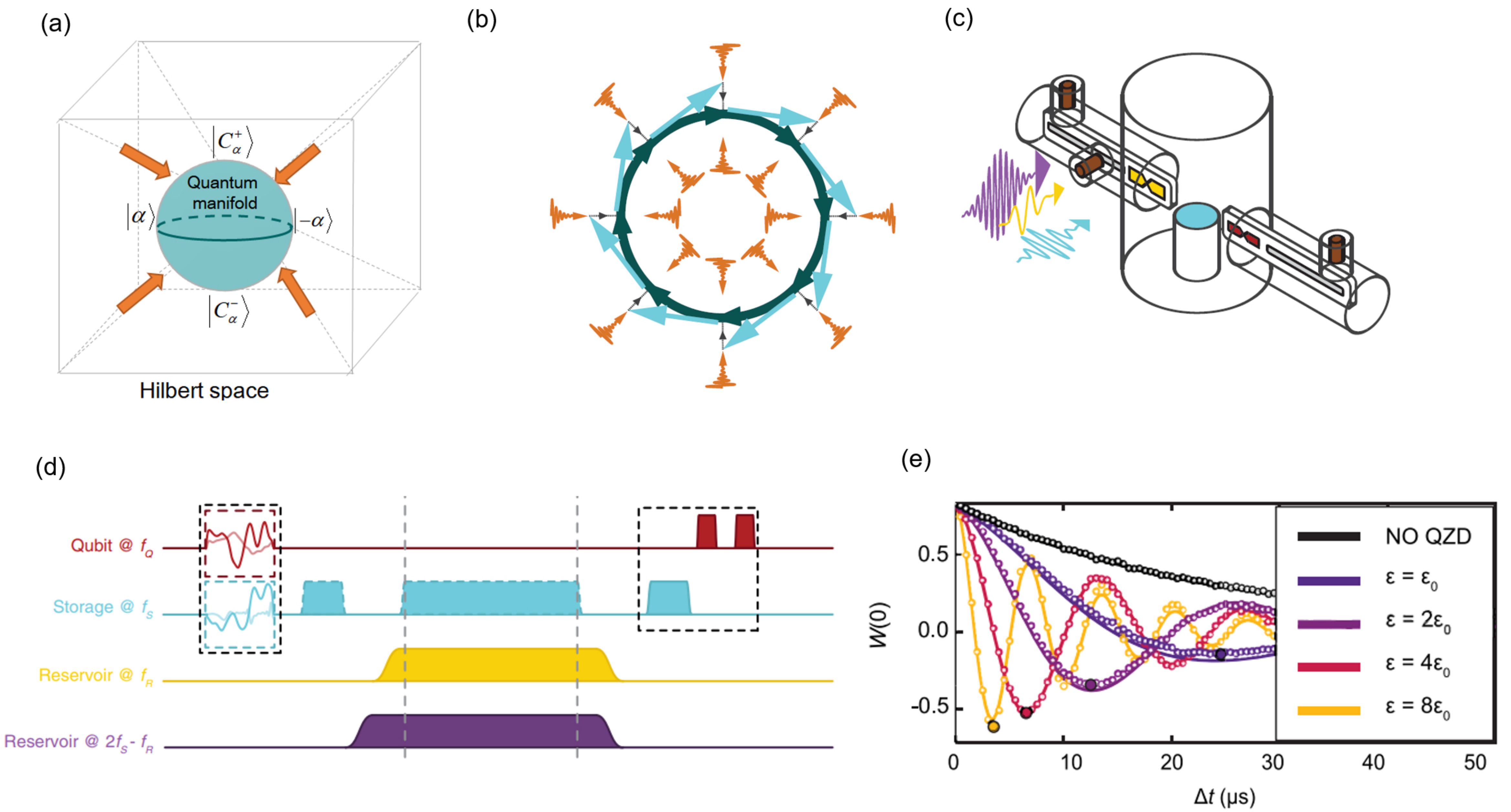}
\caption{ Formation and control of a stabilized manifold by reservoir engineering. (a) Confinement of a quantum state belonging to a large Hilbert space into a two-dimensional quantum manifold spanned by $\{|C_{\alpha}^{+}\rangle, |C_{\alpha}^{-}\rangle\}$. The cube represent a multi-dimensional Hilbert space, while the sphere represents the manifold of states. Stabilizing forces (orange arrows) direct all states toward the inner sphere without inducing any rotation in this subspace. (b) Conceptual representation of quantum Zeno dynamics in the stabilized manifold. The dark blue circle represents a cross section of the Bloch sphere of a two-state manifold in (a). The quantum Zeno dynamics corresponds to the motion along the circle. The trajectory induced by a drive in the large Hilbert space has a component both along the circle and out of it. The nonlinear dissipation and drive (orange arrows) cancels the movement outside the circle while leaving only the rotation on the circle. (c) Schematics of the experimental device. The quantum manifold is stabilized within the Hilbert space of the fundamental mode of an aluminium post storage cavity (cyan). This resonator is coupled to two Josephson junctions on sapphire (yellow for the reservoir and crimson for the transmon qubit), which are read out by stripline resonators (gray). Three couplers (brown) bring microwave drives into the system and
carry signals out of it. (d) Sequence of different drives in (c) for quantum Zeno dynamics. To make
resonant the conversion between one reservoir photon and two
storage photons, two pumps on the reservoir with frequencies $2f_{\rm S}-f_{\rm R}$ and $f_{\rm R}$ are used to create pairs of storage photons.  An additional linear displacement drive on the storage cavity induces the quantum Zeno dynamics. The drive on the transmon qubit is to initialize the storage cavity in the stabilized manifold and read-out the parity of the storage cavity. (e) Evolution of the measured parity of the storage cavity as a function of time. The initial cat states are even cat state $|C_{\alpha}^{+}\rangle$ with $|\alpha|^2=2,3,5$ (circles, squares, diamonds). The storage drive is either off (black markers) or on (colored markers) with
various strengths given in units of a chosen base strength $\epsilon_0$. Reprinted with permission from \cite{Touzard2018}.}
\label{Rabi}
\end{figure*}

The inevitable coupling of a quantum system to the reservoir generally deteriorates the coherence and coherent control of the system. However, in some cases, the system can be driven into a unitarily evolving steady subspace, which can encode and process the quantum information while being largely immune to environmental noise. This can be achieved by either reservoir engineering (designing both the system Hamiltonian and the coupling to the reservoir) or Hamiltonian engineering (designing only the system Hamiltonian), which are both called driven-dissipative control in this paper.
In this section, we will discuss the formation and control of stabilized manifolds of Schr\"{o}dinger cat states in cavity bosonic modes with both approaches. %\cite{Mirrahimi2014,Leghtas2015,Touzard2018,Albert2019}.

\subsection{Reservoir engineering}
Reservoir engineering is a powerful technique to realize steady state or subspace in condensed matter physics and quantum information processing \cite{Poyatos1996,Diehl2008,Verstraete2009}, since the steady state is often an exotic phase of matter that is difficult to stabilize in nature \cite{Diehl2008}, while the steady subspace may be used to store, protect and process quantum information \cite{Verstraete2009}.
In particular, when the quantum system is coupled to a Markovian reservoir, the time evolution of the system is governed by the Lindblad master equation \cite{Gardiner2000},
\begin{align}\label{Lind}
    \dot{\rho}=&\mathcal{L}\rho
    =-i[H,\rho]+\sum_l\mathcal{D}[F_l]\rho,
\end{align}
where the Liouvillian $\mathcal{L}$ is a superoperator on the system, $H$ is the Hamiltonian of the system including the driving term, $\mathcal{D}[F_l]\rho=2F_l\rho F_l^{\dagger}-F_l^{\dagger}F_l\rho-\rho F_l^{\dagger}F_l$ is the Lindbladian dissipator with $F_l$ being the dissipation-inducing jump operator that can depend on a parameter. The Markovian reservoir engineering refers to the design of the system Hamiltonian $H$ and the jump operators $\{F_l\}$, so that a stabilized manifold consisting of multiple steady states \cite{Albert2014,Albert2016b,Mirrahimi2014} is formed to encode quantum information and even allow QEC.

\subsubsection{Stabilized manifold with quantum information}
\textit{Single-mode two-photon process}.  Consider that a single cavity mode is driven by an external field such that it can only absorb photons in pairs, and the energy decay of the mode also happens in pairs of photons, then the Lindbladian master equation describing such a two-photon driven-dissipative process is
\begin{align}\label{single-2ph}
    \dot{\rho}=&[\epsilon_{2}a^{\dagger 2}-\epsilon_{2}^{*}a^{2},\rho]
    +\mathcal{D}[\sqrt{\kappa_{2}}a^2]\rho \nonumber \\
    =&\mathcal{D}[\sqrt{\kappa_{2}}(a^2-\alpha^2)]\rho,
\end{align}
where $\epsilon_{2}$ and $\kappa_{2}$ are the driving amplitude and decay rate, respectively. The second line of the above equation shows that the driven-dissipative dynamics can be described by a single Lindbladian dissipator $\mathcal{D}[\sqrt{\kappa_{2}}(a^2-\alpha^2)]$ with $\alpha=\sqrt{2\epsilon_{2}/\kappa_{2}}$. The stabilized manifold is determined by $\mathcal{D}[\sqrt{\kappa_{2}}(a^2-\alpha^2)]\rho=0$, and any state satisfying $a^2|\xi\rangle=\alpha^2|\xi\rangle$ or $a|\xi\rangle=\pm\alpha|\xi\rangle$ is in this manifold. Such a stabilized manifold also forms a decoherence-free subspace \cite{Lidar2013}. The stabilized manifold for two-photon process is the two-dimensional Hilbert space spanned by two coherent states $\{|\alpha\rangle, |-\alpha\rangle\}$ (Fig. \ref{Rabi}a). For any initial state $\rho(0)$, the cavity mode asymptotically converges to some pure or mixed state $\rho(\infty)$ in such a stabilized manifold. For example, if the initial state is the vacuum state $|0\rangle$ or the single-photon Fock state $|1\rangle$, the asymptotic state is the pure even ($|C_{\alpha}^{+}\rangle$) or odd ($|C_{\alpha}^{-}\rangle$) Schr\"{o}dinger cat state with
\begin{align}\label{}
    |C_{\alpha}^{\pm}\rangle=\mathcal{N}_2(|\alpha\rangle\pm|-\alpha\rangle),
\end{align}
where $\mathcal{N}_2$ is a normalization constant.

The logical qubit can be encoded into the even-odd Schr\"{o}dinger cat states $\{|C_{\alpha}^{+}\rangle, |C_{\alpha}^{-}\rangle\}$ (with large $\alpha$ so that $|\langle C_{\alpha}^{+}|C_{\alpha}^{-}\rangle|\approx0$) \cite{Mirrahimi2014}. A qubit encoded in such a way is called the \textit{dissipative-cat qubit}. For such a logical qubit, the dephasing error $\mathcal{D}[\sqrt{\kappa_{\phi}}a^{\dagger}a]$ can be largely suppressed when $\kappa_{\phi}\ll\kappa_{2}$, while the single photon loss error $\mathcal{D}[\sqrt{\kappa_{1}}a]$ causes a bit-flip error and therefore cannot be suppressed by the two-photon process. Experimentally Leghtas \textit{et al.} \cite{Leghtas2015} first successfully confined the quantum states of a superconducting cavity to the stabilized manifold spanned by the even-odd cat states.

\textit{Single-mode four-photon process}.
The four-photon process is described by letting both the absorption from the driving field and the energy decay into the bath happen through quadruples of photons,
\begin{align}\label{single-4ph}
    \dot{\rho}=&[\epsilon_{4}a^{\dagger 4}-\epsilon_{4}^{*}a^{4},\rho]
    +\mathcal{D}[\sqrt{\kappa_{4}}a^4]\rho \nonumber \\
    =&\mathcal{D}[\sqrt{\kappa_{4}}(a^4-\beta^4)]\rho.
\end{align}
The stabilized manifold is the four-dimensional Hilbert space spanned by $\{|\pm\beta\rangle,|\pm i\beta\rangle\}$ with $\beta=(2\epsilon_{4}/\kappa_{4})^{1/4}$. When the cavity mode starts at initial Fock states $|0\rangle, |1\rangle, |2\rangle, |3\rangle$, it asymptotically converges to the pure states
\begin{align}\label{}
    &|C_{\beta}^{(\rm 0mod4)}\rangle=\mathcal{N}_4(|C_{\beta}^{+}\rangle+|C_{i\beta}^{+}\rangle), \nonumber \\
    &|C_{\beta}^{(\rm 1mod4)}\rangle=\mathcal{N}_4(|C_{\beta}^{-}\rangle-i|C_{i\beta}^{-}\rangle), \nonumber \\
    &|C_{\beta}^{(\rm 2mod4)}\rangle=\mathcal{N}_4(|C_{\beta}^{+}\rangle-|C_{i\beta}^{+}\rangle), \nonumber \\
    &|C_{\beta}^{(\rm 3mod4)}\rangle=\mathcal{N}_4(|C_{\beta}^{-}\rangle+i|C_{i\beta}^{-}\rangle),
\end{align}
which form the four-component subspace of the Schr\"{o}dinger cat states.

To suppress the single photo loss error, which is usually the dominant error channel of the cavity modes, we can use the encoding scheme that can track the single-photon jump event and perform QEC. This can be achieved by encoding the qubit into the logical subspace spanned by the two cat states $\{|C_{\alpha}^{\rm 0mod4}\rangle,|C_{\alpha}^{\rm 2mod4}\rangle\}$ with even photon number parity. Then a single photon loss changes the photon number parity from even to odd. The photon number parity of the cavity mode can be monitored in a QND manner by a Ramsey experiment on an ancilla transmon qubit dispersively coupled to the cavity.

\textit{Single-mode $d$-photon process}. The two-photon and four-photon processes can be generalized to $d$-photon processes ($d=2,4,6,\cdots$ being an even integer) with
\begin{align}\label{single-dph}
    \dot{\rho}=&[\epsilon_{d}a^{\dagger d}-\epsilon_{d}^{*}a^{d},\rho]
    +\mathcal{D}[\sqrt{\kappa_{d}}a^{d}]\rho \nonumber \\
    =&\mathcal{D}[\sqrt{\kappa_{d}}(a^{d}-\gamma^{d})]\rho,
\end{align}
with $\gamma=(2\epsilon_{d}/\kappa_{d})^{1/d}$. The stabilized manifold is the $d$-dimensional Hilbert spanned by $\{|\gamma \lambda_{\nu}\rangle\}$ with $\lambda_{\nu}=\exp(2i\pi\nu/d)$ ($\nu=0,1,\cdots,d-1$), which are $d$ coherent states lying equidistantly in the phase space. The asymptotic states or cat code are $d$ different superpositions of such $d$ coherent states $\{|C_{\gamma}^{\mu{\rm mod}d}\rangle\}$ ($\mu=0,1,\cdots,d-1$) with
\begin{align}\label{}
    |C_{\gamma}^{\mu{\rm mod}d}\rangle=\mathcal{N}_{d}\sum_{\nu=0}^{2d-1}\lambda_{\nu}^{-\mu}|\gamma \lambda_{\nu}\rangle,
\end{align}
which is a superposition of $\mu$mod$d$ Fock states. The $d$-dimensional Hilbert cat space can be divided into $d/2$ subspaces labeled by $s=0,1,\cdots,d/2-1$, where the $s$-subspace is spanned by two states $\{|C_{\gamma}^{s{\rm mod}d}\rangle,|C_{\gamma}^{(s+d/2){\rm mod}d}\rangle\}$ and may encode a logical qubit \cite{Li2017,Bergmann2016}. After losing $k$ photons, the $s$ subspace is mapped to the $s-k$ subspace. Hence we can distinguish up to $d/2-1$ photon losses without destroying the encoded logical states by projectively measuring the excitation number mod $d/2$ (called the ``$\mathbb{Z}_{d/2}$ measurement"). We can also encode a qudit into the $d/2$-dimensional subspace $\{|C_{\gamma}^{0{\rm mod}d}\rangle,|C_{\gamma}^{2{\rm mod}d}\rangle,\cdots, |C_{\gamma}^{(d-2){\rm mod}d}\rangle\}$ that can correct a single photon loss error.

\textit{Multimode processes}.
The driven-dissipative processes can be extended from a single cavity mode to a two modes with operators $\{a,b\}$ \cite{Albert2019}. Suppose that both modes simultaneously absorb energy from the driving field and release energy to the bath through pairs of photons,
\begin{align}\label{multi-4ph}
    \dot{\rho}=&[\epsilon_{\rm p4}b^{\dagger 2}a^{\dagger 2}-\epsilon_{\rm p4}^{*}a^{2}b^{2},\rho]
    +\mathcal{D}[\sqrt{\kappa_{\rm p4}}a^2 b^2]\rho \nonumber \\
    =&\mathcal{D}[\kappa_{\rm p4}(a^2 b^2-\delta^4)]\rho.
\end{align}
The stabilized manifold is spanned by the pair-coherent/Barut-Girardello states \cite{Perelomov1986}. Quantum information encoded in a subspace of such a manifold is immune to the dephasing errors in both modes. Most interestingly, arbitrary photon loss errors in either mode can be corrected by continuously monitoring the photon number difference between the two modes. The two-mode generalization above can also be extended to the multimode case, with the
additional advantage of being able to correct for higher-weight products of losses or for photon
losses and gains at the same time \cite{Albert2019}.

\subsubsection{Quantum gates by quantum Zeno dynamics}

We have shown that the logical qubit encoded in the stabilized manifold can be dynamically protected from the photo loss and dephasing errors and therefore act a good quantum memory. It is also possible to perform universal gates on such a logical qubit. The arbitrary rotations around $x$-axis of a single qubit and the two-qubit entangling gate
can be generated by quantum Zeno dynamics.

When a quantum system is frequently measured to determine whether it is in the initial state, the system will always stay in the initial state, which is called the quantum Zeno effect \cite{Misra1977}. But if frequent measurements are performed to see if it is in a multi-dimensional subspace, the system is not freezed but evolves according to an effective Hamiltonian obtained by projecting the initial Hamiltonian into the measurement subspace. Such dynamics are called Quantum Zeno dynamics \cite{Facchi2002}. The driven-dissipative processes act as a continuous measurement on the quantum system to see if it is in the multi-dimensional stabilized manifold, so if we apply another driving Hamiltonian $H$, the effective driving Hamiltonian is $H_{\rm eff}=P_CHP_C$ with $P_C$ being the projector onto the stabilized manifold  (Fig. \ref{Rabi}b).

For the two-photon process with the logical qubit $\{|C_{\alpha}^{+}\rangle, |C_{\alpha}^{-}\rangle\}$, we may apply a linear drive on the oscillator, $H_{x}=\epsilon_x(a+a^{\dagger})$. The two-photon process acts as a continuous measurement which projects the driving Hamiltonian onto an effective $x$-axis rotation Hamiltonian in the qubit space,
\begin{align}\label{Heff2}
    P_{C}H_{x}P_{C}=\Omega_{x}X,
\end{align}
where $P_{C}=|C_{\alpha}^{+}\rangle\langle C_{\alpha}^{+}|+|C_{\alpha}^{-}\rangle\langle C_{\alpha}^{-}|$, $X=|C_{\alpha}^{+}\rangle\langle C_{\alpha}^{-}|+|C_{\alpha}^{-}\rangle\langle C_{\alpha}^{+}|$ and $\Omega_{x}=\epsilon_{x}(\alpha+\alpha^{*})$. One can see that a population transfer between the even cat state $|C_{\alpha}^{+}\rangle$ and the odd cat state $|C_{\alpha}^{-}\rangle$ is enabled by a resonant single-photon drive on the system (Fig. \ref{Rabi}d-f). Recently Touzard \textit{et al.} \cite{Touzard2018} have experimentally observed such coherent oscillations between the even and odd cat states by tuning the desired dissipation rate (two-photon loss rate $\kappa_2$) to be 2 orders of magnitude larger than the undesired dissipation rate (single-photon loss rate $\kappa_1$)  (Fig. \ref{Rabi}c).

For the four-photon process with the logical qubit $\{|C_{\alpha}^{\rm 0mod4}\rangle, |C_{\alpha}^{\rm 2mod4}\rangle\}$, the population transfer between two logical states needs a two-photon drive $H_{x2}=\epsilon_{x2} (a^2+a^{\dagger2})$ with the projected Hamiltonian in the stabilized manifold as
\begin{align}\label{}
     P_{C'} H_{x2}P_{C'}=\Omega_{x2}(X_{02}+X_{13}),
\end{align}
where $P_{C'}=\sum_{i=0}^3|C_{\alpha}^{i\rm mod4}\rangle\langle C_{\alpha}^{i\rm mod4}|$, $X_{ij}=|C_{\alpha}^{i\rm mod4}\rangle\langle C_{\alpha}^{j\rm mod4}|+|C_{\alpha}^{j\rm mod4}\rangle\langle C_{\alpha}^{i\rm mod4}|$ and $\Omega_{x2}=\epsilon_{x2}(\beta^2+\beta^{*2})$. The above effective Hamiltonian have two driving components: one acting on the qubit subspace to drive the Rabi oscillation between $|C_{\alpha}^{\rm 0mod4}\rangle$ and $|C_{\alpha}^{\rm 2mod4}\rangle\}$, and the other one acting on the remaining subspace to drive the Rabi oscillation between $|C_{\alpha}^{\rm 1mod4}\rangle$ and $|C_{\alpha}^{\rm 3mod4}\rangle\}$ with the same driving amplitude. Such a gate has the additional advantage of being error-transparent to single-photon loss error, since in the stabilized manifold the single photon loss operator commutes with the effective Hamiltonian and therefore can be detected/corrected at the end of the gate without compromising the encoded quantum information \cite{Vy2013,Kapit2018}. For a general $d$-photon processes to a qubit, the $x$-axis rotation Hamiltonian is $H_{x,d}=\epsilon_{x,d} (a^{d}+a^{\dagger d})$.

In addition to the $x$-axis single-qubit gates, the two-qubit entangling gates can be realized by applying appropriate driving fields. To complete the set of universal gate, we may turn off the driven-dissipative control and apply a Kerr Hamiltonian to implement single-qubit $\pi/2$-rotation around the $z$ axis \cite{Mirrahimi2014}.
The universal control of the qubits encoded in single modes can be extended to those encoded in multiple modes \cite{Albert2019}. For example, an arbitrary $x$-axis rotation of the qubit encoded in double modes can be realized by the drive $H_{\rm p}=\epsilon_{x,\rm p}(ab+b^{\dagger}a^{\dagger})$. %Here we add that robust universal control of cat qubits can also achieved by a Kerr nonlinearity with two-photon driving and non-adiabatic evolutions \cite{Puri2017,Goto2016}.

\subsection{Hamiltonian engineering}

\begin{figure*}
\includegraphics[width=7in]{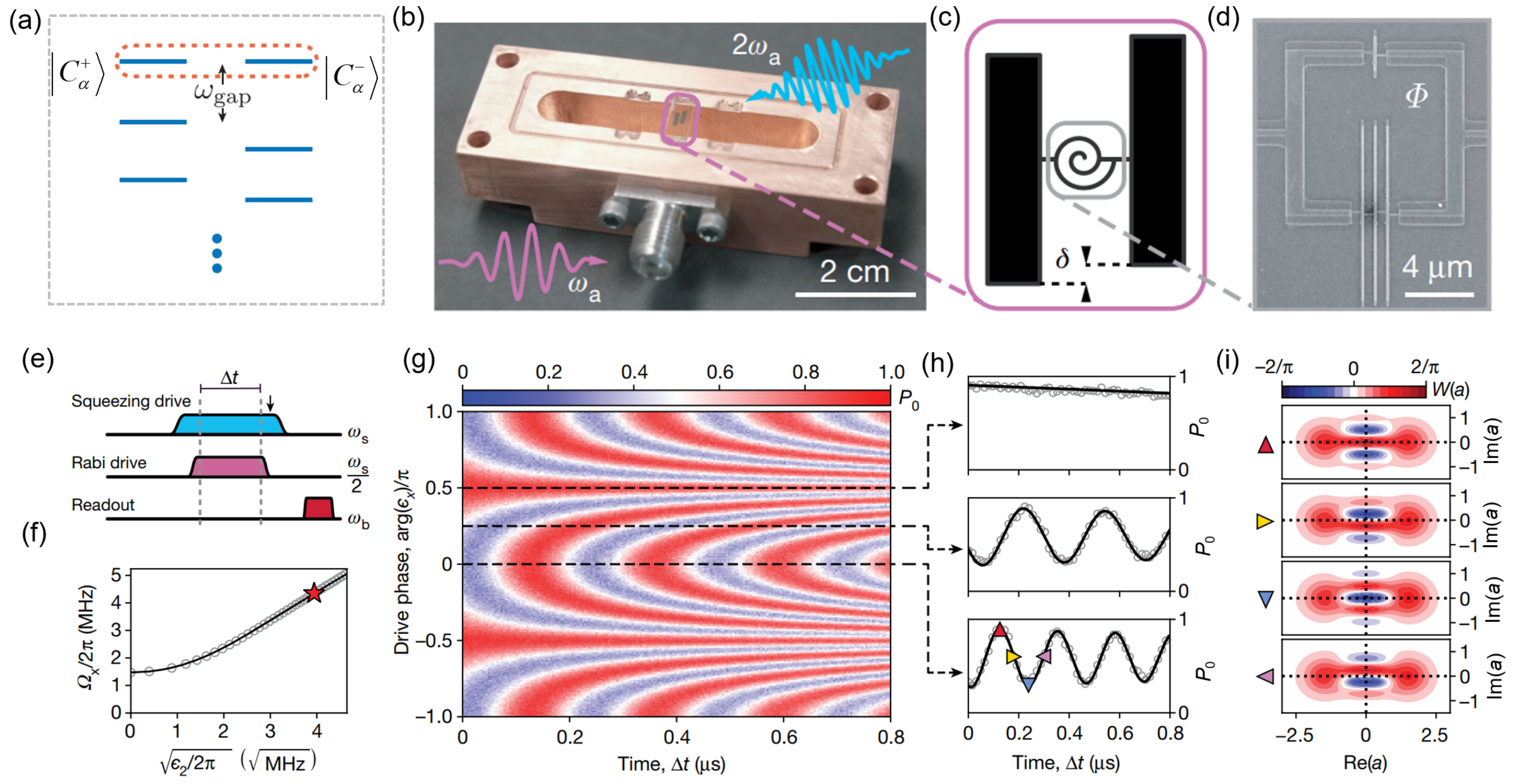}
\caption{ Formation and control of a stabilized manifold by Hamiltonian engineering. (a) Illustration of the eigenspectrum of a Kerr-nonlinear resonator with a squeezing drive. The even-odd cat states $|C_{\alpha}^{\pm}\rangle$ are two eigenstates with a large energy gap from the other eigenstates. (b) Photograph of the nonlinear resonator (purple frame) inside the copper section of the readout cavity. Also represented are the
$x$-axis rotation drive ($\omega_{a}$) and the squeezing-generation drive ($2\omega_{a}$). Here $\omega_{a}$ denotes the resonator frequency and is equivalent to $\omega_{\rm C}$ in Eq. (\ref{disp}). (c) Schematic of the nonlinear resonator with pad offset $\delta$ to set the dispersive coupling to the readout cavity and spiral symbol representing the nonlinear inductor (SNAIL element). (d) Scanning electron micrograph of the SNAIL element consisting of four Josephson junctions in a loop threaded by an external magnetic flux. (e) Pulse sequence for initialization (to $|C_{\alpha}^{+}\rangle$), Rabi oscillation and readout of the Kerr-cat qubit. Here $\omega_{s}=2\omega_{a}$, and $\omega_{b}$ is the frequency of the readout cavity. (f) Dependence of the Rabi frequency on $\sqrt{\varepsilon_2}$. (g) Dependence of the experimentally measured Rabi oscillations on evolution time $\Delta t$ and on the phase of the Rabi drive arg($\varepsilon_2$). (h) Cuts of (g) for the three Rabi-drive phases indicated by dashed lines. (i) Simulated Wigner function of the oscillator density matrix corresponding to the symbols in the bottom panel of (h). Reprinted with permission from \cite{Grimm2020}.}
\label{Rabi-Kerr}
\end{figure*}

Apart from reservoir engineering, it is also possible to form and process a stabilized manifold by only Hamiltonian engineering. The stabilized manifold can be chosen to be a degenerate eigenspace of the system with a designed Hamiltonian, which is typically decoupled to the remaining eigenspace by a large energy gap and therefore can be protected from specific system errors (Fig. \ref{Rabi-Kerr}a).

\subsubsection{Formation of the Kerr-cat qubit}
Consider the Hamiltonian of a Kerr-non\-linear resonator under the application of a single-mode squeezing drive \cite{Puri2017,Goto2016}, written in a frame rotating at the resonator frequency $\omega_{\rm C}$,
\begin{align}
{H}_{\rm Kerr}&=-K a^{\dag 2}a^2+(\epsilon_2 a^{\dag 2}+\epsilon_2^*a^2), \nonumber \\
&=-K\left(a^{\dag 2}-\alpha^{*2}\right)\left(a^{2}-\alpha^2 \right)+\frac{|\epsilon_2|^2}{K}.
\label{H0_2}
\end{align}
Here $K$ is the strength of the nonlinearity and $\alpha=\sqrt{\epsilon_2/K}$.
The second line makes it clear that the even- and odd-parity cat states $|C_{\alpha}^{\pm}\rangle$ are the degenerate eigenstates of this Hamiltonian~\cite{Puri2017,Goto2016}. This cat subspace is separated from the rest of Hilbert space by a gap $\omega_{\rm gap}\propto4K|\alpha|^2$ \cite{Puri2017}. A qubit encoded in such a way is called the {\it Kerr-cat qubit}. Observe that as the strength of the two-photon drive decreases, that is $|\epsilon_2|\rightarrow 0$ and hence $|\alpha|\rightarrow 0$, the states $|C_{\alpha}^{\pm}\rangle$ continuously approach the vacuum and single-photon Fock state, respectively. In fact, in this limit the Kerr-cat qubit is essentially the well-known transmon which encodes a ``Fock qubit" in the two photon-number states: vacuum and single-photon Fock state. It follows that, an initially undriven Kerr-nonlinear resonator ($\equiv$ Fock-qubit) prepared in vacuum or single-photon Fock state will respectively evolve to the states $|C_{\alpha}^{+}\rangle$ or $|C_{\alpha}^{-}\rangle$ as the amplitude of the squeezing drive is increased adiabatically. For the adiabatic condition to be satisfied, the rate of change of the two-photon drive must be slower than the minimum energy gap, $\dot{|\epsilon_2}(t)|/|\epsilon_2(t)| \ll2K$. So typically a large Kerr-nonlinearity results in faster cat state. Nevertheless, it is possible to apply counter-adiabatic two-photon drive to go faster than the adiabatic condition would allow~\cite{Puri2017}.

Like the case of a dissipative-cat qubit [Eq. (\ref{single-2ph})], the probability of a bit-flip error (e.g., due to frequency fluctuations $\mathcal{D}[\sqrt{\kappa_\phi}\hat{a}^\dag\hat{a}]$) is exponentially suppressed compared to a phase-flip error (for example due to single photon loss $\mathcal{D}[\sqrt{\kappa}\hat{a}]$) in the Kerr-cat qubit as well. While the dissipative-cat qubit is protected against bit-flip errors by a decoherence-free subspace enabled by engineered dissipation \cite{Lescanne2020,Guillaud2019,Chamberland2020}, the Kerr-cat qubit is protected from such errors by the underlying eigenspace structure of the two-photon driven Kerr-non\-linear resonator \cite{Puri2019,Puri2019b,Grimm2020}. Interestingly, the Kerr- and dissipative-cat qubit realizations are completely compatible with each other and have complementary properties~\cite{Puri2017,Puri2019,Puri2019b}. The inherent nonlinearity of the Kerr-cat mode provides the ability to implement fast, high-fidelity gates. It also naturally provides the ability to parametrically engineer two-photon dissipation, which can be subsequently used for autonomous correction of possible leakage errors~\cite{Puri2017,Puri2019,Puri2019b}. Recently, the adiabatic preparation of Kerr-cat was experimentally demonstrated~\cite{Grimm2020} and the asymmetry in the bit- and phase-flip errors was also confirmed. Figure~\ref{Rabi-Kerr}b-d shows the device of the superconducting setup for realization of the Kerr-cat qubit \cite{Grimm2020}.

\subsubsection{Quantum gates for the Kerr-cat qubit}
Selective control of the dynamics of the Kerr-cat qubit in the two-dimensional subspace $\{|C_{\alpha}^{+}\rangle, C_{\alpha}^{-}\rangle\}$ is possible because of the energy gap $\omega_\mathrm{gap}$ separating the qubit subspace from the rest of the Hilbert space.

Consider a coherent microwave tone applied to the resonator at the resonator's resonance frequency $\omega_\mathrm{\rm C}$.
In the rotating frame, the resulting Hamiltonian is ${H}_1={H}_{\rm Kerr}+\varepsilon_x a^\dag +\varepsilon_x^*a$. The ${a}^\dag$ term can cause transitions outside the cat-subspace. However, these transitions are off-resonant and in the limit $|\varepsilon_x|\ll \omega_\mathrm{gap}$, leakage out of the qubit subspace can be neglected. Similar to Eq. (\ref{Heff2}), the effective Hamiltonian in the qubit subspace is
\begin{align}
    {P}_C{H}_{1}{P}_C=\Omega_x{X}-\Omega_y{Y}
\label{zgate}
\end{align}
where ${P}_C$, $X$ are the same as those defined in Eq. (\ref{Heff2}), $Y=i|C_{\alpha}^{-}\rangle\langle C_{\alpha}^{+}|-i|C_{\alpha}^{+}\rangle\langle C_{\alpha}^{-}|$, $\Omega_x=\alpha(\varepsilon_x+\varepsilon_x^*)\left( r^{-1}+ r\right)/2$, $\Omega_\mathrm{y}= i\alpha(\varepsilon_x^*-\varepsilon_x)\left( r^{-1}- r\right)/2$ and $r={\sqrt{1-e^{-2\alpha^2}}}/\sqrt{1+e^{-2\alpha^2}}$.
Consequently, a resonant coherent microwave drive applied in phase with the squeezing drive ($\varepsilon_x=\varepsilon_x^*$) causes Rabi-oscillations around the $x$-axis and hence implements a $X(\theta)=\exp(i\theta X/2)$ operation ~\cite{Puri2017,Puri2019,Puri2019b}, where $\theta=\Omega_xT$ with $T$ being the evolution time. Since $r-r^{-1}\sim 2 e^{-2\alpha^2}$ in the limit of large $\alpha$, the Rabi oscillations around $y$-axis is exponentially suppressed with $\alpha^2$. The Rabi oscillations of the Kerr-cat qubit were demonstrated in a recent experiment~\cite{Grimm2020}, as shown in Fig.~\ref{Rabi-Kerr}e-i. Readout of the Kerr-cat qubit can be realized by coupling the Kerr-cat cavity to a line resonator with a beam-splitter interaction followed by a homodyne measurement of the line resonator \cite{Grimm2020}.
Furthermore, it follows from Eq.~\eqref{zgate} that a resonant beam-splitter interaction, generated parametrically between two driven nonlinear resonators, leads to an Ising coupling and realizes a $XX(\theta)=\exp{(i\theta X_1 X_2})$ gate~\cite{Puri2017,Puri2019,Puri2019b}.

The Kerr-cat qubit has an asymmetric noise channel such that $Y$ and ${Z}$ errors are exponentially suppressed. This asymmetry illustrates that the qubit couples to the environment predominantly along the $x$-axis, while coupling along the $y$ and $z$-axis is suppressed. This coupling asymmetry, also evident from Eq.~\eqref{zgate}, results from the Hilbert-space structure of the Hamiltonian of the driven Kerr-nonlinear resonator. Consequently, in order to allow coupling to the $z$-axis, it becomes necessary to turn-off the two-photon pump. When this pump is turned off, the cat-qubit freely evolves under the Kerr-nonlinearity and a $Z(\pi/2)$ gate is realized after a duration $\pi/2K$ ~\cite{Yurke1988,Kirchmair2013,Grimm2020}. After finishing the operation, the two-photon pump can be turned on again in order to recover the Kerr-cat qubit. It is important to note that unlike the $X(\theta)$ and $XX(\theta)$ gates, a $Z(\pi/2)$ rotation propagates a $X$ error as a linear combination of $X$ and $Y$ errors and consequently destroys the underlying asymmetric noise structure of the qubit~\cite{Puri2019b}.\\

Remarkably, recent theory shows that it is possible to realize a two-qubit, controlled-Z (CZ) gate without turning off the two-photon drive and thereby preserving the structure of the noise bias (termed as CX gate in ~\cite{Puri2019b} due to the different bases adopted there). The ability to implement a bias-preserving CZ gate makes the Kerr-cat qubits desirable for efficient quantum error correction. Moreover, the CZ gate can be implemented with parametric drives and four-wave mixing via the inherent Kerr-nonlinearity in the cat-qubit mode, which is very convenient to realize as no additional coupling elements are required.

\section{Holonomic quantum control} \label{holonomic}

In the last section, we have shown that through Markovian reservoir engineering, the Lindbladian dynamics can be designed to support a  mult-dimensional stabilized manifold or decoherence-free subspace to encode the quantum information without suffering dissipation, and it is also possible to realize univeral control of the states in the stabilized manifold with the aid of the quantum Zeno dynamics. In this section, we will show that the universal control of the states in the stabilized manifold can be achieved by an alternative method - holonomic quantum control.

In holonomic quantum computation (HQC) \cite{Lidar2013,Zanardi1999,Pachos1999}, the qubit states undergo adiabatic closed-loop parallel transport in parameter spaces, acquiring Abelian Berry phases \cite{Berry1984} or non-Abelian adiabatic quantum holonomies \cite{Wilczek1984} to achieve noise-resistant universal computation.
Recently Albert \textit{et al}. \cite{Albert2016a,Albert2016b} introduced the idea of HQC to Markovian reservoir engineering and found that universal computation of a quantum system consisting of superpositions of well-separated coherent states of single or multiple harmonic oscillators can be achieved by three families of adiabatic holonomic gates, including the loop gates, collision gates for single oscillator mode and controlled-phase gates for multiple harmonic oscillators. Below we will briefly introduce the first two family of gates.

Consider the following Lindbladian for a single oscillator,
\begin{align}\label{}
    \dot{\rho}=\mathcal{D}\left[\kappa\prod_{\nu=0}^{d-1}\left(a-\alpha_{\nu}(t)\right)\right]\rho.
\end{align}
which is a generalization of Eq.(\ref{single-dph}) that supports the stabilized manifold spanned by a set of coherent states $\{|\alpha_0(t)\rangle, \cdots, |\alpha_{d-1}(t)\rangle\}$. Note that different from the constant parameters $\{\alpha_{\nu}\}$ in the last section, $\{\alpha_{\nu}(t)\}$ here is time-dependent and can be tuned as external parameters. Then by adiabatically changing $\{\alpha_{\nu}(t)\}$ through closed paths in phase space, the stable coherent states $\{|\alpha_{\nu}(t)\rangle\}$ also undergo the same adiabatic evolutions.

\begin{figure}
\includegraphics[width=3.2in]{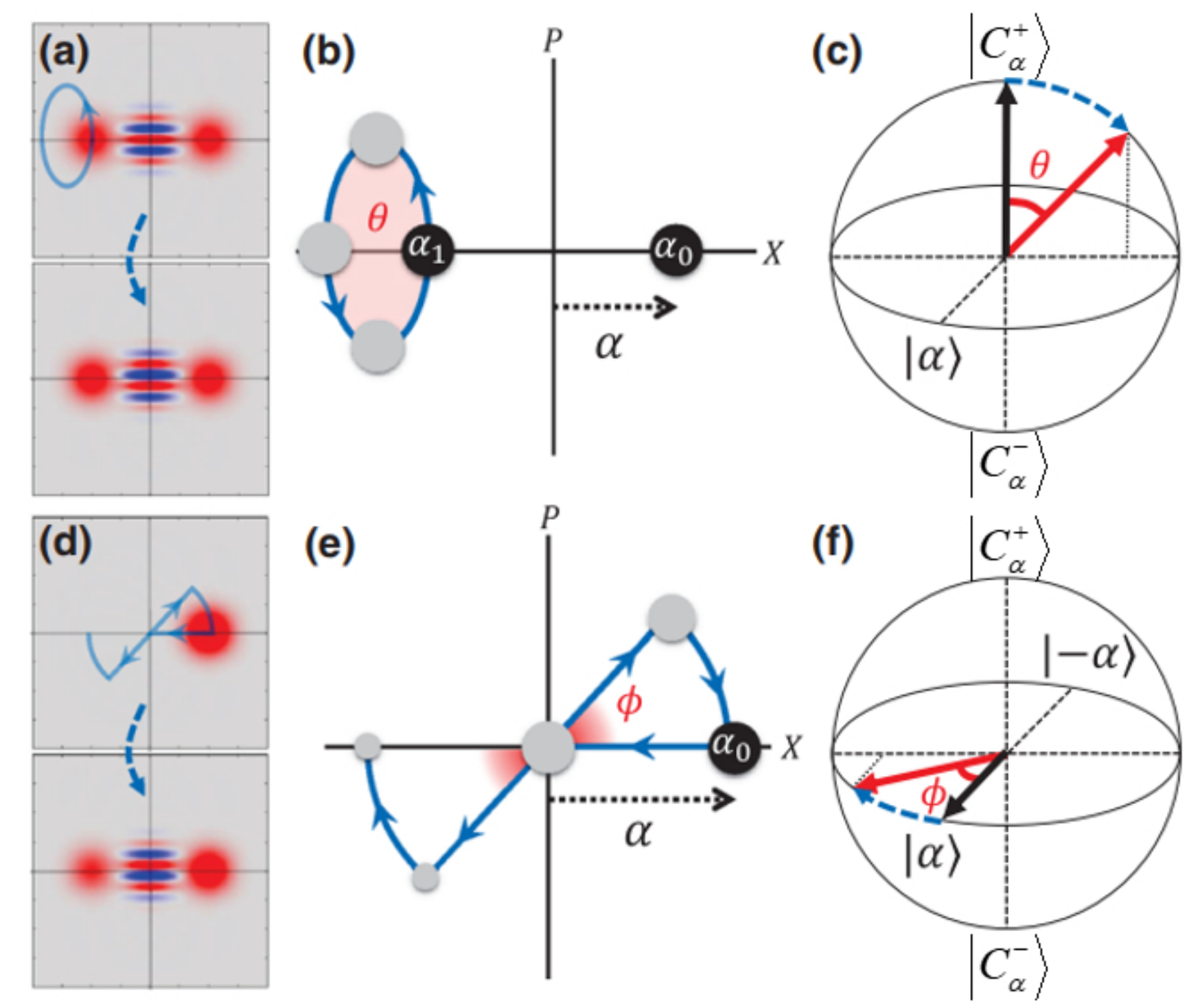}
\caption{ Holonomic gates for the logical subspace spanned by $\{|C_{\alpha}^{+}\rangle, |C_{\alpha}^{-}\rangle\}$. (a) Wigner function sketch of the state before (top) and after (bottom) a loop gate acting on $|-\alpha\rangle$, depicting the path of $|-\alpha\rangle$ during the gate (blue) and a shift in the fringes between. (b) Phase space diagram for the loop gate with $X=\langle a+a^{\dagger}\rangle/2$ and $P=-i\langle a-a^{\dagger}\rangle/2$. The parameter $\alpha_1(t)$ is varied along a closed path (blue) of area $A$, and the state $|-\alpha\rangle$ gains a phase $\theta=2A$ relative to $|\alpha\rangle$. (c) Effective Bloch sphere of the cat qubit $|\pm\alpha\rangle$ depicting the rotation caused by the loop gate. The black arrow depicts the initial state while the red arrow is the state after application of the gate. The dotted blue arrow does not represent the path traveled since the states leave the logical space during the gate. (d)-(f) Analogous descriptions of the collision gate, which consists of reducing $\alpha$ to 0, driving back to ¦Á$\alpha e^{i\phi}$, and rotating back to $\alpha$. Reprinted with permission from \cite{Albert2016a}.
.}
\label{holo}
\end{figure}

\subsection{Loop gates}
One type of the holonomic control is the loop gate, which can accumulate tunable relative Berry phases over superpositions of stabilized coherent states. First consider the simple case with $d=2$ (e.g., the single-mode two-photon process), the steady state space is $\{|\alpha_0(t)\rangle, |\alpha_1(t)\rangle\}$ with $\alpha_0(0)=-\alpha_1(0)=\alpha$  (Fig. \ref{holo}a). This stabilized manifold holds the even-odd cat qubit $\{|C_{\alpha}^{+}\rangle, |C_{\alpha}^{-}\rangle\}$. Suppose that $\alpha_1(t)$ undergoes an adiabatic variation through a closed path while $\alpha_0(t)$ is kept constant and well separated from $\alpha_1(t)$  (Fig. \ref{holo}a, b), the state $|\alpha_1(t)\rangle$ will accumulate a Berry phase $\theta=2A$ with $A$ being the area enclosed by the closed path. Such an operation is called a loop gate with implemented unitary
\begin{align}\label{}
    U_{\rm loop}&=e^{i\theta/2}\exp[-i\theta(|\alpha\rangle\langle\alpha|-|-\alpha\rangle\langle-\alpha|)/2] \nonumber \\
    &=e^{i\theta/2}\exp[-i\theta(|C_{\alpha}^{+}\rangle\langle C_{\alpha}^{-}|+|C_{\alpha}^{-}\rangle\langle C_{\alpha}^{+}|)/2],
\end{align}
which performs an $x$-axis rotation for the even-odd cat qubit (Fig. \ref{holo}c).
For the general case with an arbitrary $d$, the loop gates consist of an adiabatic evolution of $\alpha_{\nu}(t)$ around a closed path isolated from all the other $\alpha_{\nu'}(t)$ ($\nu'\neq\nu$).

\subsection{Collision gates}
The other type of the holonomic control is the collision gate, which can coherently convert the population of a stabilized coherent state to another. To get the idea of collision gates, notice that there are two distinct parameter regimes for the even-odd cat cat states: $\alpha\gg1$ and $\alpha\ll1$. In the regime $\alpha\gg1$, the cat states $|C_{\alpha}^{\pm}\rangle$ are well separated and nearly orthogonal. However, in the regime $\alpha\ll1$, the cat states are reduced to the Fock states with $|C_{\alpha}^{+}\rangle\rightarrow|0\rangle$ and $|C_{\alpha}^{-}\rangle\rightarrow|1\rangle$, so a bosonic rotation $R_{\phi}=\exp(i\phi a^{\dagger}a)$ will make the two Fock states $|0\rangle$ and $|1\rangle$ accumulate a relative phase $\phi$. If we start with the even-odd cat qubit $|C_{\alpha}^{\pm}\rangle$ with large $\alpha$, first reduce $\alpha$ to $0$, then apply bosonic rotation $R_{\phi}$ and final drive back from $0$ to $\alpha$, the net result is that $|C_{\alpha}^{+}\rangle$ and  $|C_{\alpha}^{-}\rangle$ accumulate a relative phase $\phi$ with the implemented unitary
\begin{align}\label{}
    U_{\rm coll}&=e^{i\phi/2}\exp[-i\phi(|C_{\alpha}^{+}\rangle\langle C_{\alpha}^{+}|-|C_{\alpha}^{-}\rangle\langle C_{\alpha}^{-}|)/2] \nonumber \\
    &=e^{i\phi/2}\exp[-i\phi(|\alpha\rangle\langle -\alpha|+|\alpha\rangle\langle -\alpha|)/2],
\end{align}
which performs a $z$-axis rotation for the even-odd cat qubit  (Fig. \ref{holo}f).
Denote the nonunitary driving from $0$ to $\alpha e^{i\phi}$ as $S_{\phi}$, then the collision gate can also be represented as $S_0R_{\phi}S_0^{-1}=R_{\phi}(R_{\phi}^{\dagger}S_0R_{\phi})S_0^{-1}=R_{\phi}S_{\phi}S_0^{-1}$. So an equivalent construction of the collision gate is reducing $\alpha$ to 0, driving back to $\alpha e^{i\phi}$ and rotating back to $\alpha$ (Fig. \ref{holo}d, e). The generalization of the collision gate to an arbitrary $d$ is straightforward: start with the $\{|\alpha \lambda_{\nu}\rangle\}_{\nu=0}^{d-1}$  configuration with $\lambda_{\nu}=e^{i2\pi\nu/d}$ and large enough $\alpha$, tune $\alpha$ to zero (or close to zero), pump
back to a different phase $\alpha e^{i\phi}$, and rotate back to the
initial configuration. In the cat state basis with $\alpha\ll1$, $|C_{\alpha}^{\mu{\rm mod}d}\rangle\rightarrow|\mu\rangle$ will gain a phase proportional to its mean photon number \cite{Albert2016a}.

Besides the adiabatic HQC approaches above, it is also possible to implement nonadiabatic HQC based on shortcuts-to-adiabatic (STA) dynamics \cite{Guery2019}. Recent theory shows that in the ultrastrong and deep-strong coupling regimes of the Rabi model \cite{Kockum2019}, STA methods can generate arbitrary nonclassical bosonic states and induce fast nonadiabatic gates in tens of nanoseconds \cite{Chen2021a,Chen2021b}.

\section{Multi-mode quantum control} \label{multimode}

\begin{figure*}
\includegraphics[width=5.5in]{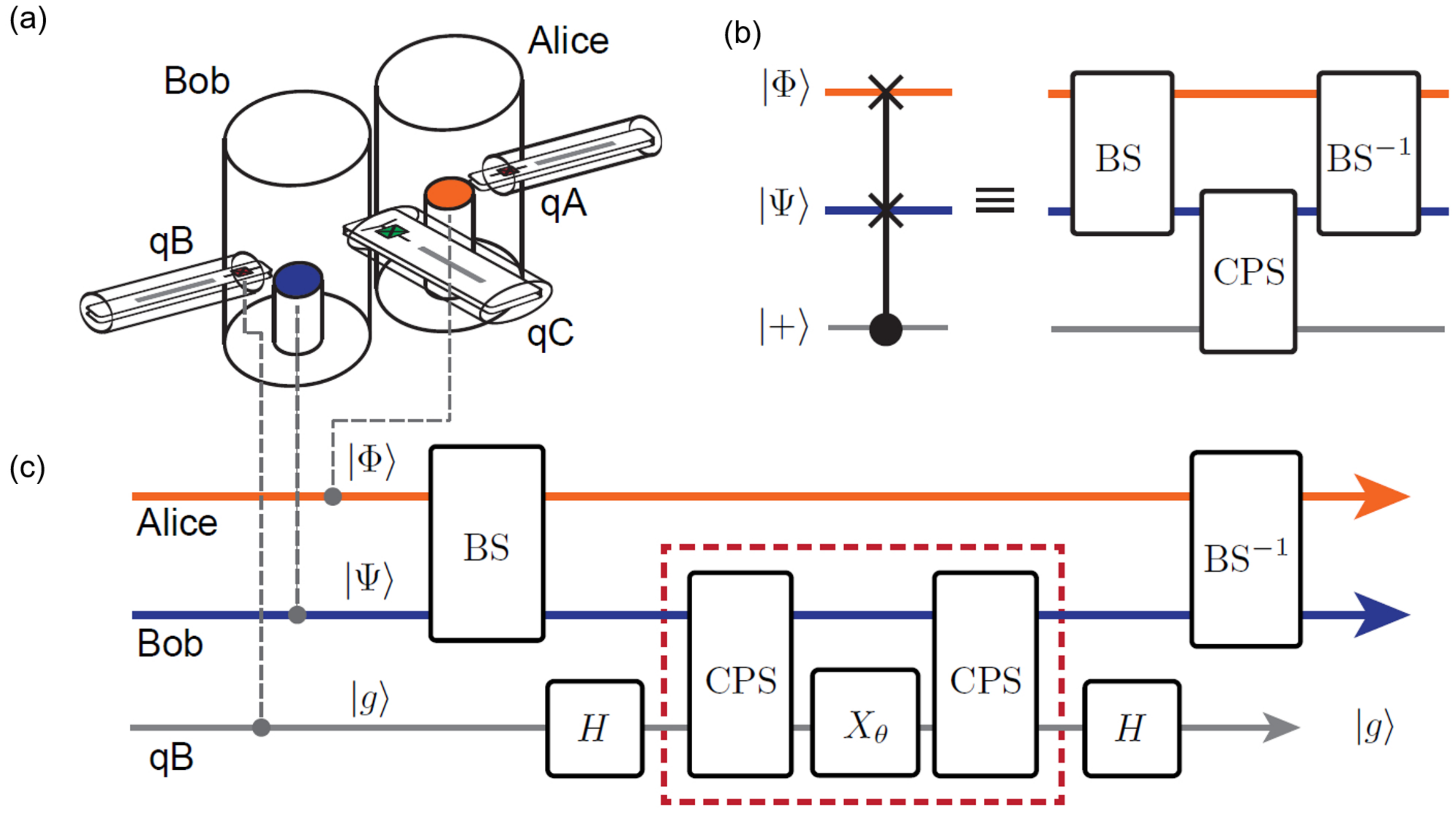}
\caption{ Design of eSWAP gate with circuit QED. (a) Schematic drawing of the three-dimensional circuit QED system used to realize the quantum Fredkin gate and eSWAP operations between two bosonic modes Alice and Bob. (b) Decomposition of
the Fredkin gate into two 50:50 beam-splitters and a controlled phase shift (CPS). The CPS for one cavity mode is described by the unitary $U_{\rm CPS}=|g\rangle\langle g|\otimes I_{\rm C}+|e\rangle\langle e|\otimes e^{i\pi a^{\dagger}a}$, which can be realized through the dispersive coupling between the cavity mode and an ancilla transmon. (c) Quantum circuit to realize the eSWAP unitary between two bosonic modes controlled by an ancilla transmon. Reprinted with permission from \cite{Gao2019}.}
\label{eSWAP}
\end{figure*}

\begin{figure*}
\includegraphics[width=7in]{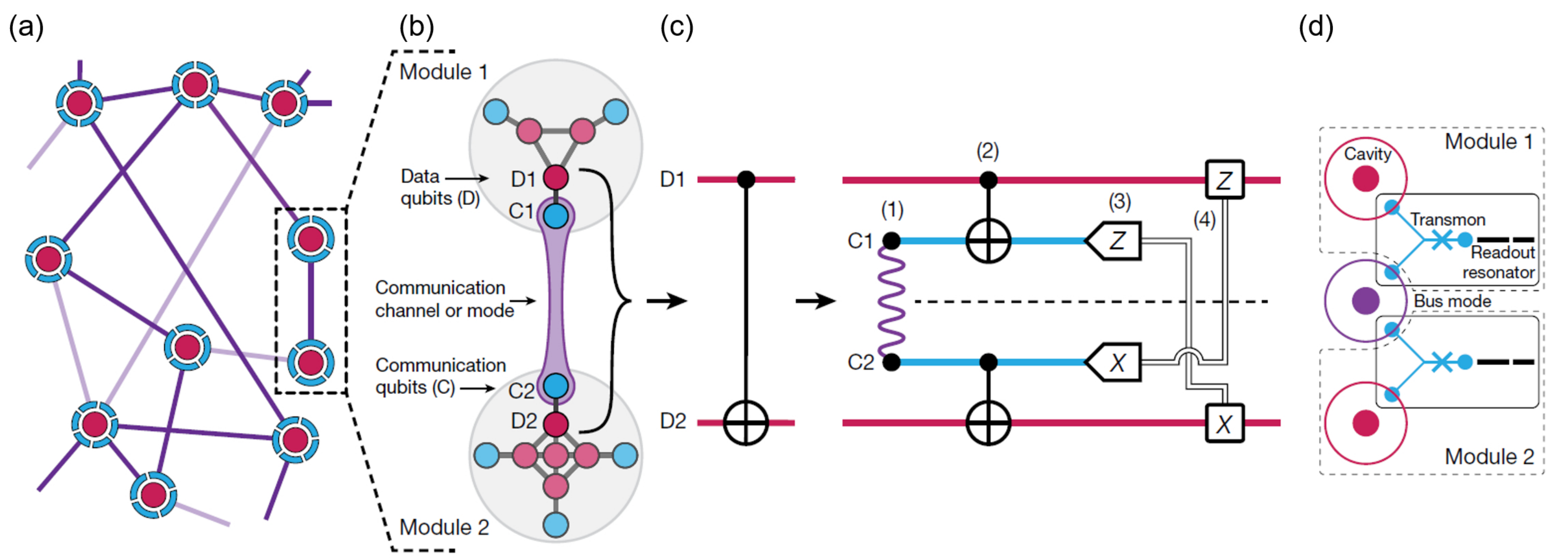}
\caption{ Modular architecture and teleported CNOT gate with circuit QED. (a) Schematic drawing of the modular quantum architecture for quantum computation. Modules are represented as nodes of a quantum network, and coupling between modules is generated through reconfigurable communication channels that can be enabled (dark purple lines) or disabled (light purple lines). (b) Each module acts a small quantum processor that is capable of high-fidelity operations among data qubits (magenta) and communication qubits (cyan). Here two modules are shown with data qubit D1, D2 and communication qubits C1, C2. (c) Quantum circuit for teleported CNOT circuit between D1 and D2. (d) Experimental realization of modular structure in circuit QED. The data qubit is realized by a high-$Q$ cavity (magenta), a communication qubit by a transmon qubit (cyan) and a low-$Q$ readout resonator (black), and the communication channel by another cavity with a bus mode. Reprinted with permission from \cite{Chou2018}.}
\label{tele-CNOT}
\end{figure*}

In all the sections above, we have concentrated on the quantum control of single bosonic mode, although sometimes we briefly mention the extension to the multi-mode control. In this section, we will focus on the quantum control of multiple bosonic modes, mainly how to entangle two bosonic modes, which is a prerequisite for universal quantum computation. Recent experimental advances in circuit QED to entangle cavity modes include preparation of two-mode cat state \cite{Wang2016}, two-mode W state \cite{Chakram2020} and  three-mode W state \cite{Chakram2020}, on-demand state transfer and entanglement generation \cite{Axline2018,Campagne-Ibarcq2018,Kurpiers2018,Zhong2018,Leung2018}, and the realization of CNOT gate \cite{Rosenblum2018a}, controlled-Z (CZ) gate \cite{Xu2020}, exponential-SWAP (eSWAP) gate \cite{Gao2019} and teleported CNOT gate \cite{Chou2018}. Below we will introduce the eSWAP gate and teleported CNOT gate.

\subsection{Exponential-SWAP gate}

The eSWAP gate can coherently transfer the states between two bosonic modes, regardless of the choice of encoding \cite{Lau2016}. To illustrate the eSWAP gate, we first introduce the unitary SWAP operation $S_{ij}$ between any two bosonic modes $a_i$, $a_j$, defined as $S_{ij} a_i S_{ij}^{\dagger}=a_j$ and $S_{ij} a_j S_{ij}^{\dagger}=a_i$ (the same relation for $a_i^{\dagger}$, $a_j^{\dagger}$). Applying the SWAP operation twice results in $S_{ij}^2 a_{i/j}S_{ij}^{\dagger2}=a_{i/j}$, which is the identity operation $I_{ij}$ for the two bosonic modes. The eSWAP gate is defined as the unitary propagator induced by a Hamiltonian in the form of the SWAP operation
\begin{align}\label{}
    U_{ij}(\theta)=\exp(i\theta S_{ij})=\cos\theta I_{ij}+i\sin\theta S_{ij},
\end{align}
which represents a superposition of the identity and the SWAP operation with the superposition coefficient tunable by the rotation angle $\theta$. At $\theta=\pi/2, \pi/4$, the eSWAP gate is reduced to the SWAP gate and $\sqrt{\rm SWAP}$ gate, respectively. One powerful feature of the eSWAP gate is that it can entangle two bosonic modes for any bosonic code. To see this, suppose the qubit code for the bosonic mode is $|0_L/1_L\rangle_i=f_{0/1}(a_i^{\dagger})|0\rangle$ with $f_{0/1}(a_i^{\dagger})$ being a function of $a_i^{\dagger}$ and $|0\rangle_i$ the vacuum state for the $i$th bosonic mode, and the initial state of the $i$th and $j$th bosonic modes is $|0_L\rangle_i |1_L\rangle_j$, then applying the eSWAP gate results in $U_{ij}(\theta)|0_L\rangle_i |1_L\rangle_j=\cos\theta|0_L\rangle_i |1_L\rangle_j+i\sin\theta|1_L\rangle_i |0_L\rangle_j$ [Note that $S_{ij}f_{0/1}(a_i^{\dagger})S_{ij}^{\dagger}=f_{0/1}(a_j^{\dagger})$ and $S_{ij}|0\rangle_i|0\rangle_j=|0\rangle_i|0\rangle_j$].

To implement the eSWAP operations, one can first use an ancilla qubit with states $\{|g\rangle, |e\rangle\}$ coupled to the two bosonic modes to realize the controlled-SWAP or Fredkin gate $C_{ij}=|g\rangle\langle g|\otimes I_{ij}+|e\rangle\langle e|\otimes S_{ij}$. The Fredkin gate can be decomposed as  (Fig. \ref{eSWAP}b)
\begin{align}\label{}
    C_{ij}=e^{-\frac{\pi}{4}(a_i a_j^{\dagger}-a_j a_i^{\dagger})}
    e^{i\frac{\pi}{2}|e\rangle\langle e|a_i^{\dagger}a_i}
    e^{\frac{\pi}{4}(a_i a_j^{\dagger}-a_j a_i^{\dagger})},
\end{align}
where the first and last unitaries are 50:50 beam splitters and the middle one is the controlled-phase shift (CPS) of one bosonic mode conditioned on the ancilla state. The CPS operation can be achieved by a dispersive coupling between the ancilla and the bosonic mode [Eq. (\ref{disp})].
Then the eSWAP gate can be realized as (Fig. \ref{eSWAP}c)
\begin{align}\label{}
    U_{ij}(\theta)|+\rangle|\psi\rangle_{ij}=C_{ij}X_{\theta}C_{ij}|+\rangle|\psi\rangle_{ij},
\end{align}
where $|+\rangle=(|g\rangle+|e\rangle)/\sqrt{2}$, $X_{\theta}=e^{i\theta(|g\rangle\langle e|-|e\rangle\langle g|)}$ and $|\psi\rangle_{ij}$ is the wavefunction for the two bosonic modes.

Recently Gao \textit{et al.} \cite{Gao2019} have experimentally implemented the eSWAP operations in three-dimensional (3D) circuit QED system (Fig. \ref{eSWAP}a) and demonstrated high-quality deterministic entanglement between two cavity modes with several different encodings including the Fock- and coherent-state coding schemes. As opposed to the eSWAP gate, a traditional CNOT gate between the multiphoton qubits in two cavities has also be realized by the mediation of a driven ancilla transmon, with the driving pulse obtained from GRAPE optimal control algorithm \cite{Rosenblum2018a}. Moreover, a geometric method has been utilized for realizing CZ gates between two logical qubits encoded in two cavities \cite{Xu2020}.

\subsection{Teleported CNOT gate}

A promising strategy toward scalable quantum computation is to adopt a quantum modular architecture (Fig. \ref{tele-CNOT}a), which is a distributed network of modules that communicate with one another through quantum and classical channels \cite{Kimble2008,Monroe2014}. Each module is composed of two functional subsystems (Fig. \ref{tele-CNOT}b): the data qubits that store and process quantum information and the communication qubits that mediate interactions between different modules. The intra-modular operations between the data and communication qubits are performed independently in each module so that the crosstalk and residual interactions between different modules are minimized even for a scaled-up system, while the inter-modular operations between the data qubits are enabled by distributing entanglement between communication qubits.

Due to the isolation between different modules, the multi-qubit operations between modules cannot depend on direct interactions but instead utilize quantum teleported gates \cite{Eisert2000,Duan2004,Gottesman1999,Jiang2007} that are enabled by entanglement sharing, local operations and classical communications. Consider two modules with the data qubits (D1 and D2) and communication qubits (C1 and C2), the teleported CNOT gate between D1 and D2 can be implemented by the following steps (Fig. \ref{tele-CNOT}c): (1) generation of entanglement
in the communication qubits C1 and C2, (2) local operations performed within each module entangle the
data and communication qubits, (3) measurement of C1 in the Pauli-$Z$ basis and C2 in the Pauli-$X$ basis and (4)
classical communication and feedforward operations.

Recently Chou \textit{et al.} \cite{Chou2018} have experimentally realized such a teleported CNOT gate in a deterministic way in cicuit QED. The experimental architecture consists of two modules (Fig. \ref{tele-CNOT}d). Each module consists of a high-$Q$ 3D electromagnetic cavity as the data qubit, a transmon qubit as the communication qubit and a Purcell-filtered, low-$Q$ stripline resonator for readout of the transmon qubit. The local operations on the data cavity mode in each module were realized by the optimal control pulses obtained by GRAPE method (see Sec. \ref{unitary}{\color{blue}B}). The communication channel was realized by an additional cavity mode that functions as a quantum bus coupling to both communication qubits in the two modules. With the first-level bosonic binomial quantum code \cite{Michael2016}, the teleported CONT gate was implemented deterministically with the process fidelity reaching 79\%.

\section{Summary and outlook} \label{summary}

Encoding quantum information in bosonic modes is a hardware-efficient approach to quantum computation, and universal quantum control of the bosonic modes is a crucial step towards this goal. Here we have given an extensive account of the recent advances in universal control of the bosonic modes. Although the approaches in this review were initially developed in the context of circuit QED, they can be extended to various other platforms, such as cavity QED \cite{Mabuchi2002}, trapped ions \cite{Leibfried2003}, nanophotonics \cite{Tiecke2014} and Rydberg atoms \cite{Signoles2014} in the strongly dispersive regime \cite{Schuster2007}.

We have shown that universal control of a single bosonic mode can be achieved with the aid of an ancilla qubit. The SNAP gates of a harmonic oscillator (cavity resonator) can be implemented by indirect control of a dispersively coupled ancilla (transmon qubit), and the SNAP gates combined with displacement operation are sufficient for universal control. We can even construct arbitrary quantum channels for the oscillator by QND readout of the ancilla and quantum feedback control. However, it is still an open problem to find the optimal control of the qubit-oscillator system with minimized expenditure of energy and resources \cite{Glaser2015}. Another problem with this qubit-oscillator system is that the ancilla qubit usually suffer relaxation and dephasing errors during the quantum gates and these ancilla errors may propagate to the logical qubits in the oscillator and corrupt the quantum information. We have shown recent theoretical and experimental advances in this respect, including the theoretical discovery of PI gates \cite{Ma2019} and experimental realization of FT parity measurement \cite{Rosenblum2018b} and PI SNAP gates \cite{Reinhold2019} in circuit QED.

Universal quantum control can also be achieved in some noise-resilient subspace of the bosonic modes. With the aid of reservoir engineering or Hamiltonian engineering, the bosonic modes may support some multi-dimensional decoherence-free subspace to encode quantum information. Applying appropriate drive can implement the desired unitary on this stabilized manifold allowed by quantum Zeno dynamics. Universal control in such stabilized manifold can also be achieved by holonomic quantum control, where the external parameters are tuned so that the stable states undergo some adiabatic evolutions. Recent experimental advances include the formation of stabilized manifold in two-photon process \cite{Leghtas2015,Grimm2020}, Rabi population oscillations in such a manifold \cite{Touzard2018,Grimm2020} and the formation and control of a Kerr-cat qubit \cite{Grimm2020}. However, it is still challenging to experimentally generate desired engineered dissipation that is much stronger than the undesired dissipations. Moreover, it remain unsolved to systematically extract high-order nonlinear Hamiltonian of the oscillator, in order to support high-dimensional steady state subspaces.

Apart from universal control of single bosonic modes, coupling different bosonic modes is also needed for universal quantum computation. We have introduced two approaches to entangling two bosonic modes with recent experimental realizations: the eSWAP gate independent of the bosonic encoding \cite{Gao2019} and the teleported CNOT gate for a modular architecture \cite{Chou2018}. It is interesting to further design some robust generalization of eSWAP gates that are FT to the ancilla errors and bosonic loss errors. Moreover, the teleported CNOT gate has only been realized for adjacent modules, and it will be the next milestone to demonstrate the non-local teleported gates using spatially separate modules.

\section{Acknowledgements}
We acknowledge support from the ARO (W911NF-18-1-0020, W911NF-18-1-0212), ARO MURI (W911NF-16-1-0349), AFOSR MURI (FA9550-19-1-0399), NSF (EFMA-1640959, OMA-1936118, EEC-1941583), NTT Research, the Packard Foundation (2013-39273), and the Startup Foundation of Institute of Semiconductors, Chinese Academy of Sciences (No. E0SEBB11).

\section{Author contributions}
Liang Jiang, Wenlong Ma and S. M. Girvin organized the manuscript. Wenlong Ma and Liang Jiang carried out the literature search and wrote most parts of the manuscript. Shruti Puri wrote the part in Sec. IV-B. S. M. Girvin, Michel H. Devoret and Robert J. Schoelkopf contributed to the manuscript revision. All authors contributed to the discussion.

%\bibliographystyle{apsrev4-1}
%\bibliography{Refs-Rev_Quantum_control}

\end{document}

\title{Supplementary Information for ``Quantum control of bosonic modes with superconducting circuits''}
\author{Wen-Long Ma}
%\email{wenlongma@semi.ac.cn}
\affiliation{State Key Laboratory for Superlattices and Microstructures, Institute of Semiconductors, Chinese Academy of Sciences, Beijing 100083, China}
\affiliation{Pritzker School of Molecular Engineering, University of Chicago, Illinois 60637, USA}
%\affiliation{Department of Applied Physics and Physics, Yale University, New Haven, Connecticut 06511, USA}
%\affiliation{Yale Quantum Institute, Yale University, New Haven, Connecticut 06511, USA}
\author{Shruti Puri}
\affiliation{Department of Applied Physics and Physics, Yale University, New Haven, Connecticut 06511, USA}
\affiliation{Yale Quantum Institute, Yale University, New Haven, Connecticut 06511, USA}
\author{Robert J. Schoelkopf}
\affiliation{Department of Applied Physics and Physics, Yale University, New Haven, Connecticut 06511, USA}
\affiliation{Yale Quantum Institute, Yale University, New Haven, Connecticut 06511, USA}
\author{Michel H. Devoret}
\affiliation{Department of Applied Physics and Physics, Yale University, New Haven, Connecticut 06511, USA}
\affiliation{Yale Quantum Institute, Yale University, New Haven, Connecticut 06511, USA}
\author{S. M. Girvin}
\affiliation{Department of Applied Physics and Physics, Yale University, New Haven, Connecticut 06511, USA}
\affiliation{Yale Quantum Institute, Yale University, New Haven, Connecticut 06511, USA}
\author{Liang Jiang}
\email{liang.jiang@uchicago.edu}
\affiliation{Pritzker School of Molecular Engineering, University of Chicago, Illinois 60637, USA}
%\affiliation{Department of Applied Physics and Physics, Yale University, New Haven, Connecticut 06511, USA}
%\affiliation{Yale Quantum Institute, Yale University, New Haven, Connecticut 06511, USA}

\date{\today }

\maketitle

\section{Implementation of driven-dissipative processes in circuit QED}

In this appendix, we will discuss the physical realization of reservoir engineering or driven-dissipative processes, in particular, how to construct the desired Lindbladian jump operators in Eq. (13) of the main text. We will first take the two-photon driven-dissipative process in circuit QED as an example and then provide the general recipe.

\begin{figure}
\includegraphics[width=3.0in]{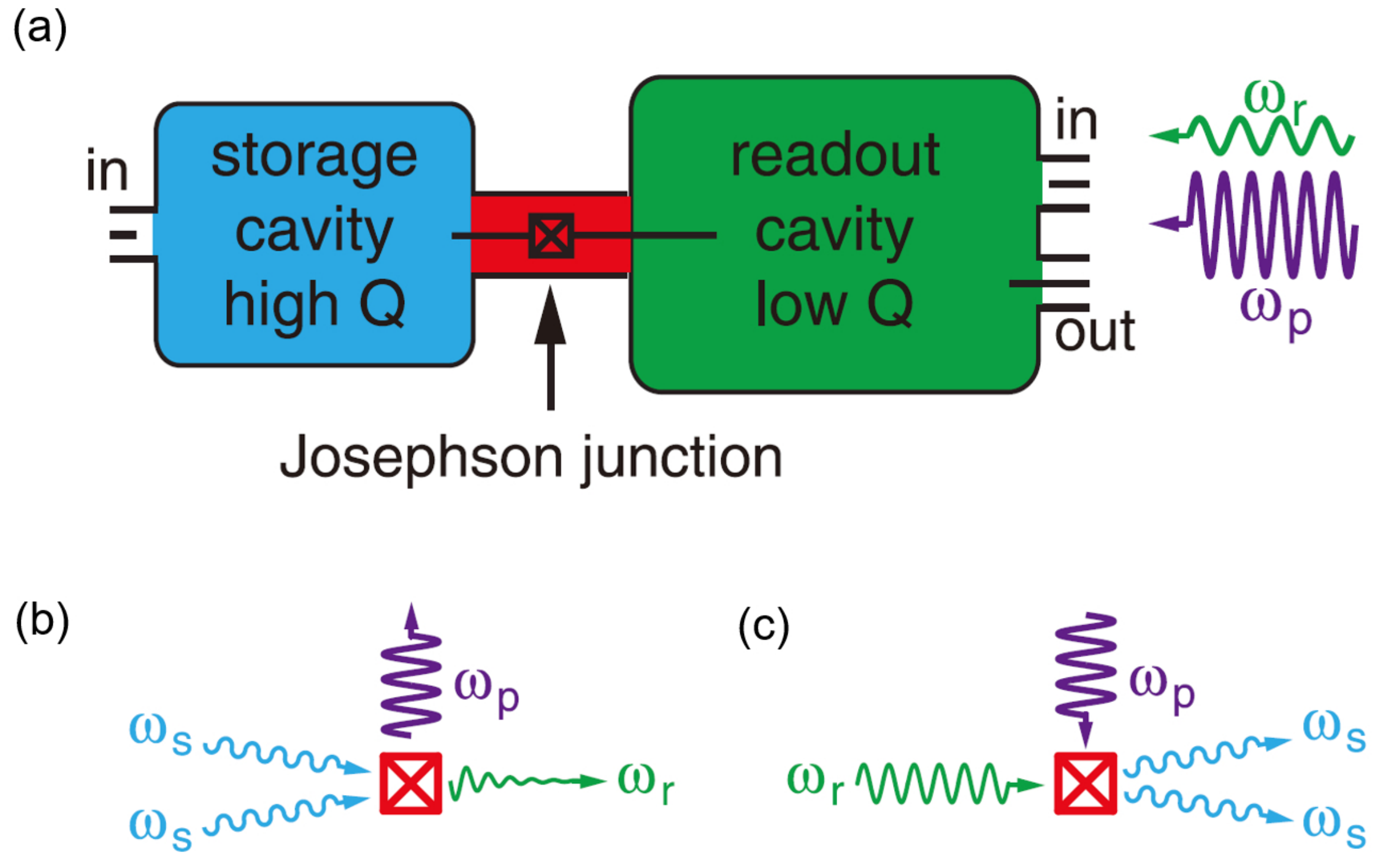}
\caption{Implementation of two-photon driven-dissipative processes in circuit QED. (a) Two superconducting cavities (a high-$Q$ storage cavity and a low-$Q$ storage cavity) are coupled through a Josephson junction. A pump and a drive microwave tones are applied to the readout cavity, creating the appropriate nonlinear interaction that generates a coherent superposition of steady states in the storage. (b) Four-wave process experienced by the storage cavity, where two photons of the storage combine and convert, stimulated by the pump tone,
into a readout photon that is irreversibly radiated away by the transmission line. (c) Reversed four-wave process as opposed to that in (b). Reprinted with permission from \cite{Leghtas2015}.  }
\label{imple}
\end{figure}

Consider a high-$Q$ storage cavity and a low-$Q$ readout cavity linked by a Josephson junction (Fig. \ref{imple}a). The Josephson junction provides a nonlinear coupling between the modes in the two cavities \cite{Nigg2012},
\begin{align}\label{}
    H_0=\sum_k \omega_k a^{\dagger}_{k}a_{k}-E_J(\cos({\Phi}/{\phi_0})+(\Phi/\phi_0)^2/2),
\end{align}
where $\Phi=\sum_k\phi_k(a_{k}+a^{\dagger}_{k})$, $E_J$ is the Josephson energy, $\phi_0=1/2e$ is the reduced superconducting flux quantum, $\phi_k$ is the standard deviation of the zero point flux fluctuation for mode $k$ with frequency $\omega_k$. Note that the bare frequencies of the cavity modes are usually shifted by the nonlinear couplings. We are concerned with only one fundamental mode in each cavity, with $a_{\rm r}/a_{\rm s}$, $\omega_{\rm r}/\omega_{\rm s}$ being the annihilation operator, the frequency of the mode for the storage/readout cavity (already accounting for the frequency shift by nonlinear couplings), respectively. Then we apply a weak resonant drive tone and a strong off-resonant pump tone on the readout cavity
\begin{align}\label{}
    H_{p}=[\epsilon_{\rm b}(t)+\epsilon_{\rm p}(t)](a_{\rm r}+a_{\rm r}^{\dagger}),
\end{align}
where $\epsilon_{\rm b}(t)=2\epsilon_{\rm b} \cos(\omega_{\rm r} t)$, and $\epsilon_{\rm p}(t)=2\epsilon_{\rm p} \cos(\omega_{\rm p} t)$ with $\omega_{\rm p}=2\omega_{\rm s}-\omega_{\rm r}$. In the rotating frame associated with $H_{\rm rs0}=\omega_{\rm s}a_{\rm s}^{\dagger}a_{\rm s}+\omega_{\rm r}a_{\rm r}^{\dagger}a_{\rm r}$ and with the rotating wave approximation, we get an effective Hamiltonian describing the storage and readout cavities,
\begin{align}\label{Hrs}
    H_{\rm sr}=&(g_2 a_{\rm s}^2 a_{\rm r}^{\dagger}+g_2^{*}a_{\rm s}^{\dagger2} a_{\rm r})
    -\epsilon_{p}(a_{\rm r}+a_{\rm r}^{\dagger})+\chi_{\rm ss}(a_{\rm s}^{\dagger}a_{\rm s})^2 \nonumber \\
    &+\chi_{\rm rr}(a_{\rm r}^{\dagger}a_{\rm r})^2
    +\chi_{\rm sr}(a_{\rm s}^{\dagger}a_{\rm s})(a_{\rm r}^{\dagger}a_{\rm r}),
\end{align}
where $\chi_{\rm ss}/\chi_{\rm rr}$ is the self-Kerr coefficient for the storage/readout cavity and $\chi_{\rm sr}$ is the cross-Kerr coefficient between the storage and readout cavity. Now let us add the single photon loss channel for mode $a_{\rm r}$ with the effective Lindbladian master equation
\begin{align}\label{Hrs}
    \dot{\rho_{\rm sr}}=-i[H_{\rm sr}, \rho_{\rm sr}]+\mathcal{D}[\sqrt{\kappa_{\rm r}}a_{\rm r}]\rho_{\rm sr}.
\end{align}
After neglecting the Kerr and cross-Kerr terms in $H_{\rm sr}$, we can adiabatically eliminate mode $a_{\rm r}$ in the above master equation \cite{Carmichael1988} and get an effective master equation only for mode $a_{\rm s}$ [Eq. (14) of the main text] with the corresponding parameters $\epsilon_2=2\epsilon_{\rm b}g_2/\kappa_{\rm b}$, $\kappa_2=4g_2^2/\kappa_{\rm b}$ and $\alpha=\sqrt{\epsilon_{\rm b}/g_2}$.

We add three points for the above adiabatic transformation: (i) the pump tone on the readout cavity (the first term in $H_{\rm sr}$) stimulates a conversion of a pair of photons in mode $a_{\rm s}$ to a single photon in mode $a_{\rm r}$, so a single photon loss in mode $a_{\rm r}$ (the dissipation term $\mathcal{D}[\sqrt{\kappa_{\rm r}}a_{\rm r}]$) through the lossy channel can result in a loss in photon pairs in mode $a_{\rm s}$ ($\mathcal{D}[\sqrt{\kappa_{2}}a_{\rm s}]$); (ii) the drive tone (the second term in $H_{\rm sr}$) injects energy into mode $a_{\rm r}$ and then a single photon in mode $a_{\rm r}$ can be converted to a pair of photons in mode and vice versa  (Fig. \ref{imple}b, c), resulting in the driving Hamiltonian $a_{\rm s}$ ($\epsilon_2a_{\rm s}^{\dagger 2}-\epsilon_2^{*}a_{\rm s})$; (iii) the self-Kerr and cross-Kerr couplings (the last three terms in $H_{\rm sr}$) are undesired terms for the driven-dissipative control, but fortunately has negligible effect on the scheme \cite{Mirrahimi2014,Leghtas2015}. Note that the above construction for single storage cavity (single-qubit gate) can be directly extended to the case of protectively coupling two storage cavities (two-qubit entangling gate), by introducing additional coupling between different storage cavity modes.

Similar to two-photon process, the four-photon process can be realized by designing the effective Hamiltonian of the storage and cavity modes as $H_{\rm sr}=(g_4 a_{\rm s}^4 a_{\rm r}^{\dagger}+g_4^{*}a_{\rm s}^{\dagger4} a_{\rm r})-\epsilon_{p}(a_{\rm r}+a_{\rm r}^{\dagger})$. It has been theoretically proposed that such highly nonlinear interactions between the storage and readout cavity modes can be implemented by an architecture of Josephson ring modulators \cite{Mirrahimi2014}, which may perfectly select the desired nonlinear interactions while avoiding other undesirable interactions.

Now we summarize the general recipe for realizing a jump operator $F$ in reservoir engineering \cite{Poyatos1996,Albert2016a}. First, the central system should be coupled to some auxiliary system, possibly by the nonlinear coupling Hamiltonian $Fc^{\dagger}+{\rm H.c.}$ with $c$ being the operator for the auxiliary system. Second, the auxiliary system suffers Markovian dissipation with $c$ being the Lindbladian jump operator. Then, if the thermal fluctuations in the auxiliary system can be neglected, one can adiabatically eliminate the auxiliary system and obtain an effective Lindbladian master equation for the central system with the desired jump operator $F$. Moreover, we may also need some additional drives on the auxiliary system to realize some other desired drives on the central system.

%\begin{acknowledgments}
%This work is supported by Hong Kong Research Grants
%Council¨CCollaborative Research Fund Grant No. CUHK4/
%CRF/12G and the Chinese University of Hong Kong Vice
%Chancellor¡¯s One-Off Discretionary Fund.
%\end{acknowledgments}

%\bibliographystyle{apsrev4-1}
%\bibliography{Refs-Rev_Quantum_control}